\documentclass[prd,aps,preprint,showpacs]{revtex4}

\usepackage{epsf}
\usepackage{pslatex}

\def\mpi2{m_\pi^2}
\def\mK2{m_K^2}

\newcommand{\bea}{\begin{eqnarray}}
\newcommand{\eea}{\end{eqnarray}}
\newcommand{\be}{\begin{equation}}
\newcommand{\ee}{\end{equation}}
\newcommand{\nn}{\nonumber}

\newcommand{\A}{{\cal A}}
\newcommand{\B}{{\cal B}}

\newcommand{\h}{ h}
\newcommand{\el}{ l}

\newcommand{\Mo}{M^{-1}\frac{\partial M}{\partial\mu}}
\newcommand{\Mt}{M^{-1}\frac{\partial^2 M}{\partial\mu^2}}
\newcommand{\Mth}{M^{-1}\frac{\partial^3 M}{\partial\mu^3}}
\newcommand{\Mf}{M^{-1}\frac{\partial^4 M}{\partial\mu^4}}
\newcommand{\Mfi}{M^{-1}\frac{\partial^5 M}{\partial\mu^5}}
\newcommand{\Ms}{M^{-1}\frac{\partial^6 M}{\partial\mu^6}}
\newcommand{\C}{{\cal C}}
\newcommand{\mz}{M^{-1}}
\newcommand{\mo}{\frac{\partial M^{-1}}{\partial\mu}}
\newcommand{\mt}{\frac{\partial^2 M^{-1}}{\partial\mu^2}}
\newcommand{\mth}{\frac{\partial^3 M^{-1}}{\partial\mu^3}}
\newcommand{\mf}{\frac{\partial^4 M^{-1}}{\partial\mu^4}}
\newcommand{\mfi}{\frac{\partial^5 M^{-1}}{\partial\mu^5}}
\newcommand{\ms}{\frac{\partial^6 M^{-1}}{\partial\mu^6}}
\newcommand{\fz}{\frac{dM}{du_0}}
\newcommand{\fo}{\frac{\partial}{\partial\mu}\frac{dM}{du_0}}
\newcommand{\ft}{\frac{\partial^2}{\partial\mu^2}\frac{dM}{du_0}}
\newcommand{\fth}{\frac{\partial^3}{\partial\mu^3}\frac{dM}{du_0}}
\newcommand{\ff}{\frac{\partial^4}{\partial\mu^4}\frac{dM}{du_0}}
\newcommand{\ffi}{\frac{\partial^5}{\partial\mu^5}\frac{dM}{du_0}}
\newcommand{\fs}{\frac{\partial^6}{\partial\mu^6}\frac{dM}{du_0}}

\newsavebox{\DERIVBOXZLM}
\savebox{\DERIVBOXZLM}[2.5em]{$\Longrightarrow\hspace{-1.5em}
\raisebox{.2ex}{*}
\hspace{-.7em}\raisebox{-.8ex}{\scriptsize lm}\hspace{.7em}$}

\begin{document}
\epsfclipon


\newcommand{\pbp}{\langle \bar \psi \psi \rangle}
\newcommand{\pbdmdup}{\left\langle \bar \psi \frac{dM}{du_0} \psi
\right\rangle}



\title{QCD thermodynamics with 2+1 flavors at nonzero chemical potential}

\author{C. Bernard}
\affiliation{Department of Physics, Washington University, St.~Louis,
MO 63130, USA}

\author{C. DeTar and L. Levkova} \affiliation{Physics Department, University of Utah,
Salt Lake City, UT 84112, USA}

\author{Steven Gottlieb}
\affiliation{Department of Physics, Indiana University, Bloomington,
IN 47405, USA}

\author{U.M. Heller}
\affiliation{American Physical Society, One Research Road, Box 9000,
Ridge, NY 11961-9000, USA}

\author{J.E. Hetrick}
\affiliation{Physics Department, University of the Pacific, Stockton, CA 95211, USA}

\author{R. Sugar}
\affiliation{Department of Physics, University of California, Santa
Barbara, CA 93106, USA}

\author{D. Toussaint}
\affiliation{Department of Physics, University of Arizona, Tucson, AZ
85721, USA}
\date{\today}

\begin{abstract}
We present results for the QCD equation of state, quark densities and susceptibilities at nonzero chemical 
potential, using 2+1 flavor asqtad ensembles with $N_t=4$. The ensembles
lie on a trajectory of constant physics for which $m_{ud}\approx0.1m_s$. The calculation
is performed using the Taylor expansion method with
terms up to sixth order in $\mu/T$.  
\end{abstract}

\pacs{12.38.Gc, 12.38.Mh, 25.75.Nq}

\maketitle

\newpage


\section{Introduction}
\label{sec:intro}

The equation of state (EOS) of QCD 
is of special interest to the interpretation of data from 
heavy-ion collision experiments and to the development of 
nuclear theory and cosmology. The EOS at zero chemical potential ($\mu=0$) 
has been extensively studied on the lattice. However, to approximate most closely
the conditions of heavy ion collision experiments (for example RHIC has $\mu\sim 15$ MeV 
\cite{BraunMunzinger:2001ip})
or of the interior of dense stars,
the inclusion of nonzero chemical potential is necessary.
Unfortunately, as is well known, inclusion of a nonzero chemical
potential makes the fermion determinant in numerical simulations
complex and straightforward Monte Carlo simulation not applicable.
Several methods have been developed to overcome or circumvent this
problem. They include the reweighting techniques
\cite{Barbour:1997ej,Fodor:2001au},
simulations with an imaginary chemical potential combined with analytical 
continuation \cite{Lombardo:1999cz,deForcrand:2002ci} or
canonical ensemble treatment \cite{Alford:1998sd}, and lastly, 
the Taylor 
expansion method \cite{Allton:2002zi,Gavai:2003mf}, which is employed here.
In this method one Taylor expands the quantities
needed for the computation of the EOS around the point $\mu=0$ where
standard Monte Carlo simulations are possible. The expansion parameter
is the ratio $\mu/T$, where T is the temperature.
To ensure fast convergence of the Taylor series, the expansion parameter should be
sufficiently small. Numerical calculations show satisfactory convergence for
$\mu/T\lesssim 1$ (see reviews \cite{Philipsen:2005mj,Schmidt:2006us}).

 In our simulations we use 2+1 flavors of improved staggered fermions. In such simulations
where the number of flavors is not equal to a
multiple of four, the so-called ``fourth root
trick'' is employed to reduce the number of ``tastes''. While this trick is still
somewhat controversial, there is a growing body of numerical 
\cite{Durr:2003xs} 
and analytic \cite{Shamir:2004zc} evidence that it leads
to the correct continuum limit.
For simulations at nonzero chemical potential the problems of rooting
are much more severe \cite{Golterman:2006rw}. However, the Taylor expansion 
method is not directly affected by this additional problem with rooting since the
coefficients in the Taylor series are calculated in the theory with zero
chemical potential. The Taylor expansion method is generally considered
reliable in regions where the studied physics quantities are analytic.  

The Taylor expansion method has been used to study the phase structure and the EOS of
two flavor QCD \cite{Allton:2002zi,Gavai:2004sd,Allton:2003vx,Allton:2005gk,Ejiri:2005uv}. 
Our work improves on the previous studies by the addition of the 
strange quark to the sea. Our calculations are performed on 2+1 flavor ensembles generated
with the $R$ algorithm \cite{Gottlieb:1987mq} and using the asqtad 
quark action \cite{Orginos:1998ue}
and a one-loop 
Symanzik improved gauge action \cite{Symanzik}.
These improved actions have small discretization errors 
of $O(\alpha_sa^2, a^4)$ and $O(\alpha_s^2a^2, a^4)$, respectively. 
This is very important since we study the $N_t=4$ case, where the
lattice spacing ($a = 1/(TN_t)$) is quite large, especially at low temperatures.
Our ensembles lie along
a trajectory of constant physics for which the ratio of the heavy quark
mass and the light quark mass is $m_{ud}/m_{s}\approx0.1$, and the heavy quark
mass itself is tuned approximately to the physical value of the 
strange quark mass. The determination of the Taylor expansion coefficients,
other than the zeroth order ones computed already previously, is necessary
only on the finite temperature ensembles (for our study $N_t=4$).
No zero-temperature subtractions are needed for them. We have determined
the contributions to the energy density, pressure and interaction measure due to
the presence of a nonzero chemical potential. We also present results for the
quark susceptibilities and densities. In addition, we have calculated the
isentropic EOS,
which is highly relevant for the heavy-ion collision experiments, where,
after thermalization, the created matter is supposed to expand without 
further increase in entropy or change in the baryon number. All the results
are obtained with the strange quark density fixed to $n_s=0$ regardless
of temperature, appropriate for the experimental conditions. This requires 
the tuning of the strange quark chemical potential along the trajectory
of constant physics.
 
\section{The Taylor expansion method}
\label{sec:theory}
In this section we give a brief description of the 
Taylor expansion method for the thermodynamic quantities
we study and as applied to the asqtad fermion formulation.
\subsection{Calculating the pressure}
The asqtad quark matrix for a given flavor with nonzero chemical potential is:
\bea
\hspace{-1cm}{M}_{l,h} &=
& {M}_{l,h}^{\rm spatial} 
+ \frac{1}{2} \eta_0(x) \left[ U_0^{(F)}(x) {\rm e}^{\mu_{l,h}}\delta_{x+\hat 0,y} -
 U_0^{(F)\dagger}(x-\hat 0) {\rm e}^{-\mu_{l,h}} \delta_{x,y+\hat 0} \right. \\\nn 
 &&\left. + U_0^{(L)}(x) {\rm e}^{3\mu_{l,h}} \delta_{x+3\hat 0,y} -
 U_0^{(L)\dagger}(x-3\hat 0) {\rm e}^{-3\mu_{l,h}} \delta_{x,y+3\hat 0} \right] ,
\eea
where $\mu_l = \mu_{ud}$ and $\mu_h=\mu_s$ are the quark chemical 
potentials in lattice units for the light ($u$ and $d$) quarks and the heavy
(strange $s$) quark, respectively. In the above
\bea
\hspace{-1cm} {M}_{l,h}^{\rm spatial} &=& am_{l,h} \delta_{x,y}
+ \sum_{k=1}^3 \frac{1}{2} \eta_k(x) \left[ U_k^{(F)}(x) \delta_{x+\hat k,y} -
 U_k^{(F)\dagger}(x-\hat k) \delta_{x,y+\hat k} \right. \\\nn 
 &&\left. + U_k^{(L)}(x) \delta_{x+3\hat k,y} -
 U_k^{(L)\dagger}(x-3\hat k) \delta_{x,y+3\hat k} \right] ,
\eea
with $m_{l,h}$ the light and strange quark masses.
The superscripts $F$ and $L$ on the
links $U_\mu$ denote the type of links, ``fat'' and ``long''; appropriate
weights and factors of the tadpole strength $u_0$ are included in
$U_\mu^{(F)}$ and $U_\mu^{(L)}$.
The partition function based on the asqtad quark matrix is
\be
{\cal Z} = \int {\cal D}U\,{\rm e}^{\frac{n_l}{4}\ln {\rm det}\,{ M}_l}
{\rm e}^{\frac{n_h}{4} \ln {\rm det}\,{ M}_h}{\rm e}^{-S_g},
\ee
where $n_l=2$ is the number of light quarks and $n_h=1$ is the number of heavy quarks. 
The pressure $p$ can be obtained from the identity
\be
\frac{p}{T^4} = \frac{\ln \cal Z}{T^3V},
\ee
where $T$ is the temperature and $V$ the spatial volume.
It can be Taylor expanded in the following manner
\begin{equation}
{p\over T^4}=
\sum_{n,m=0}^\infty c_{nm}(T) \left({\bar{\mu}_l\over T}\right)^n
\left({\bar{\mu}_h\over T}\right)^m,
\label{eq:p}
\end{equation}
where $\bar{\mu}_{l,h}$ is the nonzero chemical potential in physical units. 
Due to the CP symmetry of the partition function, only the terms with $n+m$
even are nonzero.
The expansion coefficients are defined by
\begin{equation}
c_{nm} (T)=
{1\over n!}{1\over m!}{N_t^{3}\over N_s^3}{{\partial^{n+m}\ln{\cal Z}}\over
{\partial(\mu_l N_t)^n}{\partial(\mu_h N_t)^m}}\biggr\vert_{\mu_{l,h}=0} \quad,
\label{eq:cn}
\end{equation}
with $\mu_{l,h}=a\bar{\mu}_{l,h}$ and $N_s$ and $N_t$ the spatial and temporal extents of the 
lattice. All coefficients need to be calculated 
on the finite-temperature ensembles only, except for $c_{00}(T)$. The latter is the
pressure divided by $T^4$ at $\mu_{l,h}=0$, which needs a zero-temperature
subtraction. It should be calculated by other means, such as the integral method, which we have already done
in \cite{Bernard:2006nj}.
The $c_{nm}(T)$ coefficients are 
linear combinations of observables $\A_{nm}$ and are given in Appendix~B. 
The $\A_{nm}$ observables are obtainable as linear combinations 
of various products of the operators
\bea
\label{eq:Ln}
L_n &=& \frac{n_l}{4} \frac{\partial^n \ln \det M_l}{\partial \mu_l^n} \\
\label{eq:Hn}
H_m &=& \frac{n_h}{4} \frac{\partial^m \ln \det M_h}{\partial \mu_h^m}, 
\label{eq:basic}
\eea 
evaluated at $\mu_{l,h}=0$. For the definitions and explicit forms of
the $\A_{nm}$ see Appendix~B.

Figure~\ref{fig:free} compares the cut-off effects due to the finite
temporal extent $N_t$ in the
free theory case for the coefficients $c_{00}$, $c_{20}$, $c_{40}$ and $c_{60}$ for 
three different staggered fermion actions: the standard, the Naik (asqtad) and
the p4 action.
The results for the first three coefficients are normalized to their respective
Stefan-Boltzmann (SB) values. 
The SB value for $c_{60}$ is zero (and the same holds for $c_{06}$) . 
In the SB limit, the $c_{0n}$ coefficients are, of course, equal to half of the 
SB values of $c_{n0}$ for $0<n\leq4$. All other coefficients with $n,m\neq 0$ are 
zero in the SB limit.  In the interacting  case, the coefficients which are 
zero in the SB limit can aquire non-zero values.
Figure~\ref{fig:free} shows that the asqtad action has better scaling properties than the standard 
(unimproved) staggered action at $N_t=4$, but it is clear that a study at
larger $N_t$ is important for further reduction of the discretization errors.
\begin{figure}[ht]
\begin{tabular}{ll}
  \epsfxsize=80mm
  \epsfbox{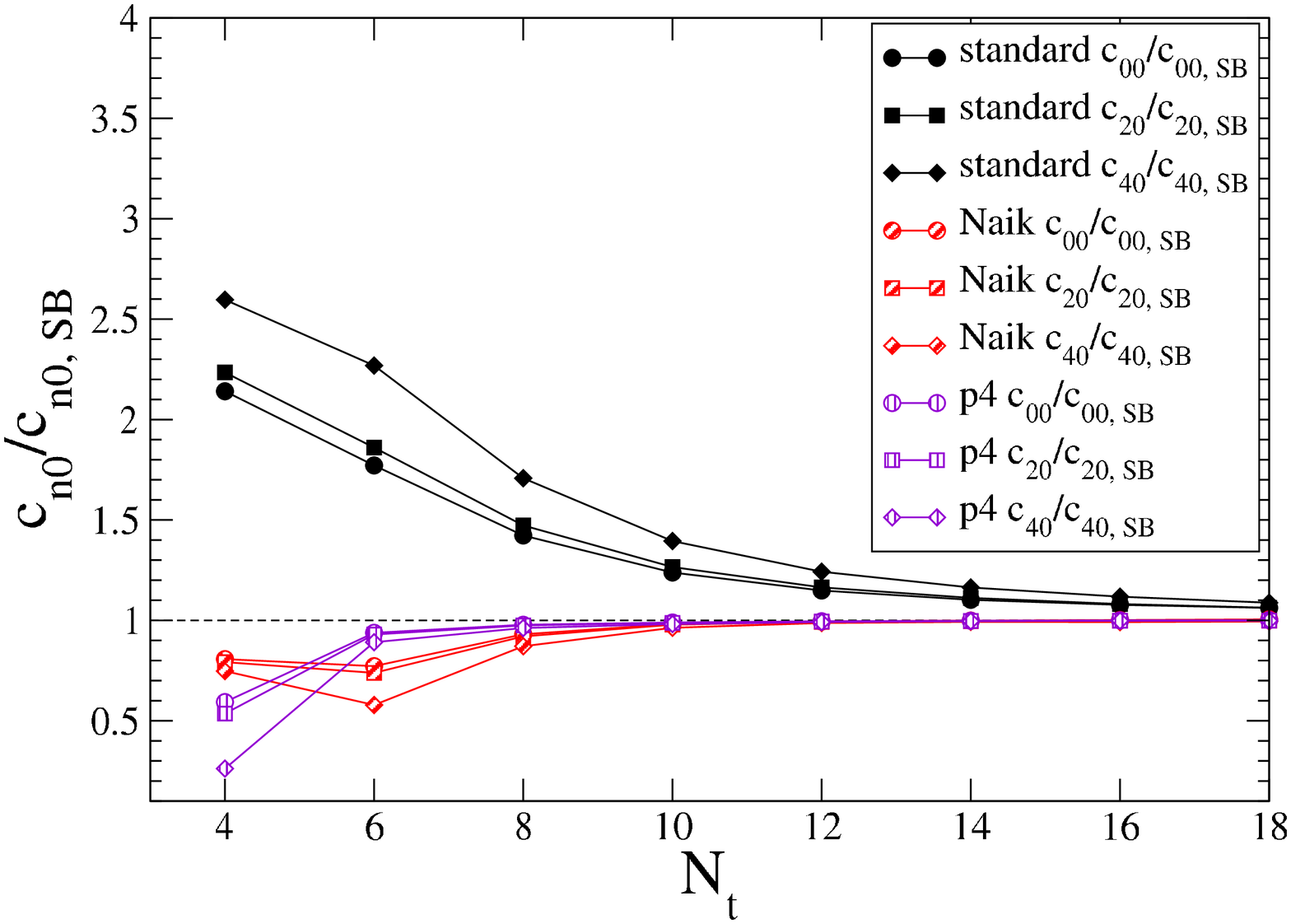}
  \epsfxsize=80mm
  \epsfbox{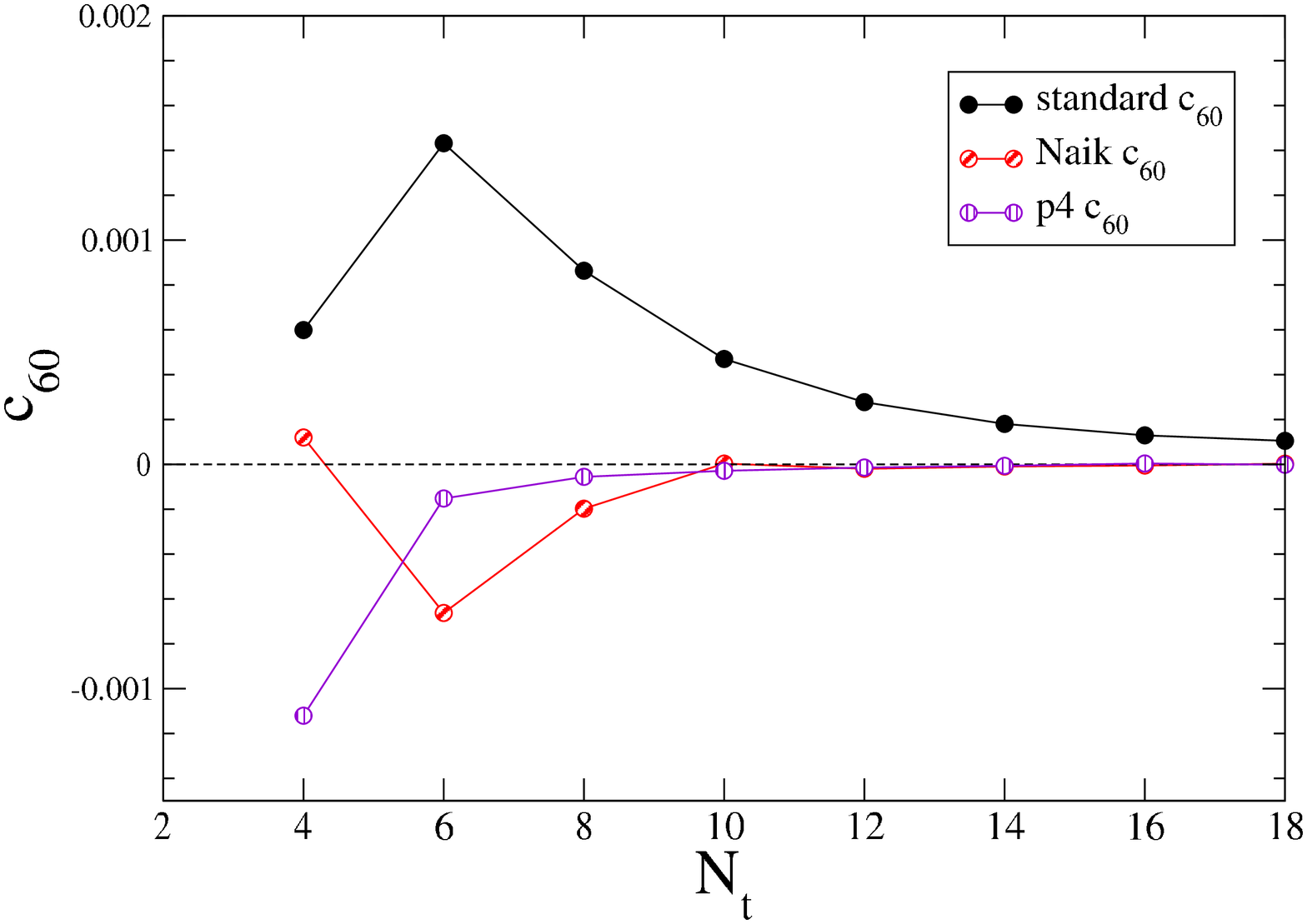}
\end{tabular}
\caption{The expansion coefficients $c_{00}$, $c_{20}$, $c_{40}$ and $c_{60}$
for the pressure in the free theory case as a function of $N_t$.}
\label{fig:free}
\end{figure}
\subsection{Calculating the interaction measure and energy density}
The interaction measure $I$ can be Taylor expanded in a manner similar to the pressure 
\be
{I\over T^4}=-{N_t^3\over N_s^3}{d\ln {\cal Z} \over d\ln a}=\sum_{n,m}^\infty b_{nm}(T)\left({\bar{\mu_l}\over T}\right)^n
\left({\bar{\mu_h}\over T}\right)^m,
\ee
where again only terms even in $n+m$ are nonzero and
\be
b_{nm}(T) = \left.-{1\over n!m!}{N_t^3\over N_s^3}{\partial^{n+m} \over \partial(\mu_l N_t)^n\partial(\mu_h N_t)^m}
\right|_{\mu_{l,h}=0}\left({d\ln {\cal Z}\over d\ln a}
\right).
\ee
The derivative with respect to $\ln a$ is taken along a trajectory of
constant physics.
The fermionic part of ${d\ln {\cal Z}\over d\ln a}$, considering the form 
of the asqtad action, is
\be
\left\langle{d\,S_f\over d\ln a}\right\rangle =\sum_{f=h,l}\frac{n_f}{4}\left[\frac{d(m_fa)}
{d\ln a}{\rm tr}
\langle M_f^{-1}\rangle +
\frac{du_0}{d\ln a}{\rm tr}\langle M_f^{-1}\frac{dM_f}{du_0}\rangle
\right].\nn
\ee
No volume normalization of the various traces is assumed in the above. 
The gauge part, taking into account the explicit form of the Symanzik gauge action,
is 
\be
\left\langle{-d\,S_g\over d\ln a}\right\rangle = \langle{\cal G}\rangle=\langle6\frac{d\beta}{d\ln a}P
+ 12\frac{d\beta_{\rm rt}}{d\ln a}R + 16\frac{d\beta_{\rm pg}}{d\ln a}C\rangle,
\ee
where $ P$, $ R$ and $C$ are the appropriate sums of the plaquette, rectangle 
and parallelogram terms, respectively (here they are not normalized to the volume).
Thus the $b_{nm}(T)$ coefficients become
\bea
\hspace{-2cm}b_{nm}(T) &=& -{1\over n!m!}{N_t^3\over N_s^3}
\sum_{f=l,h}\frac{n_f}{4}
\left[
\left.\frac{ d(m_fa)}{d\ln a}\right|_{\mu_{l,h}=0}{\rm tr}\left.{\partial^{n+m} \langle M_f^{-1}\rangle\over \partial(\mu_l N_t)^n
\partial(\mu_h N_t)^m}\right|_{\mu_{l,h}=0}\right.\nn\\
&&+
\left.\left.\frac{du_0}{d\ln a}\right|_{\mu_{l,h}=0}{\rm tr}\left.{\partial^{n+m}
\langle M_f^{-1}\frac{dM_f}{du_0}\rangle
\over \partial(\mu_l N_t)^n\partial(\mu_h N_t)^m}\right|_{\mu_{l,h}=0}   
\right]\nn\\
&& -{1\over n!m!}{N_t^3\over N_s^3}\left.{\partial^{n+m}\langle {\cal G}\rangle
\over \partial(\mu_l N_t)^n\partial(\mu_h N_t)^m}\right|_{\mu_{l,h}=0}
\eea
The explicit forms of the $b_{nm}(T)$ coefficients are more complex than those for $c_{nm}(T)$
and we save them for Appendix C. The SB limit of all $b_{nm}$ coefficients is zero.
In the presence
     of interactions their values can become different from zero. For the computation of the $b_{nm}(T)$ coefficients, in addition to the
derivatives of the fermion matrix 
and the gauge action with respect to the
chemical potentials, we have to know the derivatives of the action parameters with respect to 
$\ln a $ along the trajectory of constant physics. The latter have been determined in our previous
work on the EOS at zero chemical potential \cite{Bernard:2006nj}, along with the coefficient $b_{00}(T)$, which is the 
interaction measure divided by $T^4$ in that case.
The coefficients $c_{nm}(T)$ can be obtained from $b_{nm}(T)$ by integration along the
trajectory of constant physics. This can serve as a consistency check of the calculation.

The energy density $\varepsilon$ is simply obtained from the linear combination
\be
\frac{\varepsilon}{T^4} = \frac{I + 3p}{T^4}.
\ee
\subsection{Quark number densities and susceptibilities}
The Taylor expansion for the quark number densities can be obtained
from that for the pressure. For example, the light quark number density,
$n_{ud}$, is
\be
{n_{ud}\over T^3}=\frac{\partial}{\partial \bar{\mu}_l /T}\left(\frac{\ln \cal Z}{T^3V}\right)=
\sum_{n=1,m=0}^\infty nc_{nm}(T) \left({\bar{\mu}_l\over T}\right)^{n-1}
\left({\bar{\mu}_h\over T}\right)^m,
\ee
and the heavy one, $n_s$, is
\be
{n_{s}\over T^3}=\frac{\partial}{\partial \bar{\mu}_h /T}\left(\frac{\ln \cal Z}{T^3V}\right)=
\sum_{n=0,m=1}^\infty mc_{nm}(T) \left({\bar{\mu}_l\over T}\right)^{n}
\left({\bar{\mu}_h\over T}\right)^{m-1}.
\ee
Similarly, the quark number susceptibilities are derivatives of the 
quark number densities with respect to the chemical potentials. Thus, 
the diagonal light-light quark susceptibility becomes
\be
{\chi_{uu}\over T^2}=\frac{\partial}{\partial \bar{\mu}_l /T}\left(\frac{n_{ud}
}{T^3}\right)=
\sum_{n=2,m=0}^\infty n(n-1)c_{nm}(T) \left({\bar{\mu}_l\over T}\right)^{n-2}
\left({\bar{\mu}_h\over T}\right)^m,
\ee
and the heavy-heavy diagonal one is
\be
{\chi_{ss}\over T^2}=\frac{\partial}{\partial \bar{\mu}_h /T}\left(\frac{n_{s}
}{T^3}\right)=
\sum_{n=0,m=2}^\infty m(m-1)c_{nm}(T) \left({\bar{\mu}_l\over T}\right)^{n}
\left({\bar{\mu}_h\over T}\right)^{m-2}.
\ee 
Lastly, the mixed quark susceptibility has the form
\be
{\chi_{us}\over T^2}=\frac{\partial}{\partial \bar{\mu}_h /T}\left(\frac{n_{ud}
}{T^3}\right)=
\sum_{n=1,m=1}^\infty nmc_{nm}(T) \left({\bar{\mu}_l\over T}\right)^{n-1}
\left({\bar{\mu}_h\over T}\right)^{m-1}.
\ee


\section{Simulations}
\label{sec:sim}

The asqtad-Symanzik gauge
ensembles we use in this study have spatial volumes of $12^3$ or $16^3$ 
and $N_t=4$, and are generated using the $R$ algorithm. They are a subset of the 
ensembles in our EOS calculation at zero chemical potential \cite{Bernard:2006nj}.
The ensembles lie on an approximate trajectory of constant physics for
which $m_{ud} \approx0.1m_s$, and $m_s$ is tuned to the physical strange quark 
mass within 20\%. Along the trajectory, the $\pi$ to $\rho$ mass ratio is
$m_\pi/m_\rho\approx 0.3$.    
Table~I in \cite{Bernard:2006nj} contains the run parameters and trajectory
numbers of the ensembles 
used here. They are the ones that have the gauge coupling values  of $\beta=6.0$,
6.075, 6.1, 6.125, 6.175, 6.2, 6.225, 6.25, 6.275, 6.3, 6.35, 6.6 and 7.08. 
The last column of that table shows the lattice scale. For 
explanation of the scale setting and other simulation details 
we refer the reader to section~III of \cite{Bernard:2006nj}.
The observables that need to be measured along the trajectory 
of constant physics in order to construct the Taylor coefficients in the
expansion for the pressure are $L_n$ and $H_m$ defined by Eqs.~(\ref{eq:Ln})~and~(\ref{eq:Hn}).
For the interaction measure determination the following observables
have to be calculated in addition:
\bea
{\el}_n &=& \frac{\partial^n {\rm tr}\,M_l^{-1}}{\partial \mu_l^n},\hspace{2cm}
\h_m = \frac{\partial^m {\rm tr}\,M_h^{-1}}{\partial \mu_h^m},\\
&&\nn\\
\lambda_n &=& \frac{\partial^n {\rm tr}\, (M_{l}^{-1}\frac{dM_{h}}{du_0})}{\partial\mu_{l}^n},\hspace{1cm}
\chi_m = \frac{\partial^m {\rm tr}\, (M_{h}^{-1}\frac{dM_{h}}{du_0})}{\partial\mu_{h}^m}
\eea
and the gluonic observables $P$, $R$ and $C$.
In Appendix~C we show how they enter in the coefficients $b_{nm}(T)$. 
To sixth order in the Taylor expansion, the number of fermionic observables
($L_n$, $H_m$, $l_n$, $h_m$, $\lambda_n$, $\chi_m$) 
that need to be determined is 40. We calculate them stochastically employing
random Gaussian sources. In the region outside the phase transition or
crossover we 
use 100 sources and double that number inside the transition/crossover
region. This
ensures that we work with statistical errors dominated 
by the gauge fluctuations and not by the ones coming from the 
stochastic estimators.

The ensembles we are working with have been generated using the inexact
$R$ algorithm which introduces finite step-size errors.
In our previous study of these ensembles \cite{Bernard:2006nj} we 
measured the step-size
error in both gluonic and fermionic observables.  The error was
considerably less than 1\% in the relevant gluonic and fermionic
observables, measured on the high temperature ensembles.  For the EOS 
at zero chemical potential it is necessary to subtract the high
temperature and zero temperature values.  In the difference the effect
of the step-size error becomes somewhat more pronounced.  
The contributions to the EOS due to nonzero chemical potential,
computed here, do not require zero temperature subtractions. Thus,
based on the observations noted above, we expect any step-size errors
in these contributions to be considerably smaller than our statistical
errors.

\section{Numerical results}
\begin{figure}[h]
\epsfxsize=\textwidth
\begin{center}
\epsfbox{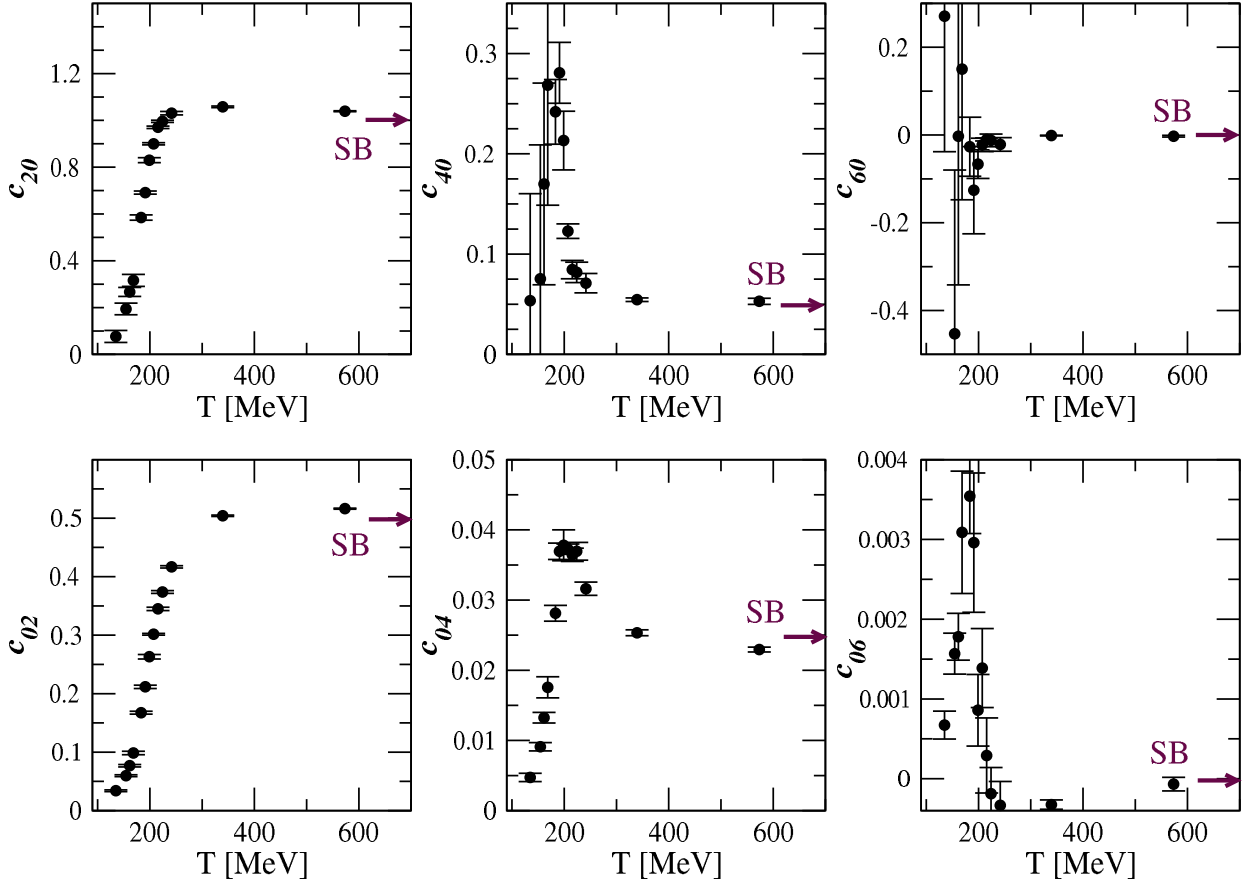}
\end{center}
\caption{Taylor expansion coefficients $c_{n0}(T)$ and $c_{0m}(T)$ for $p/T^4$.}
\label{fig:c_un}
\end{figure}

\begin{figure}[h]
\epsfxsize=\textwidth
\begin{center}
\epsfbox{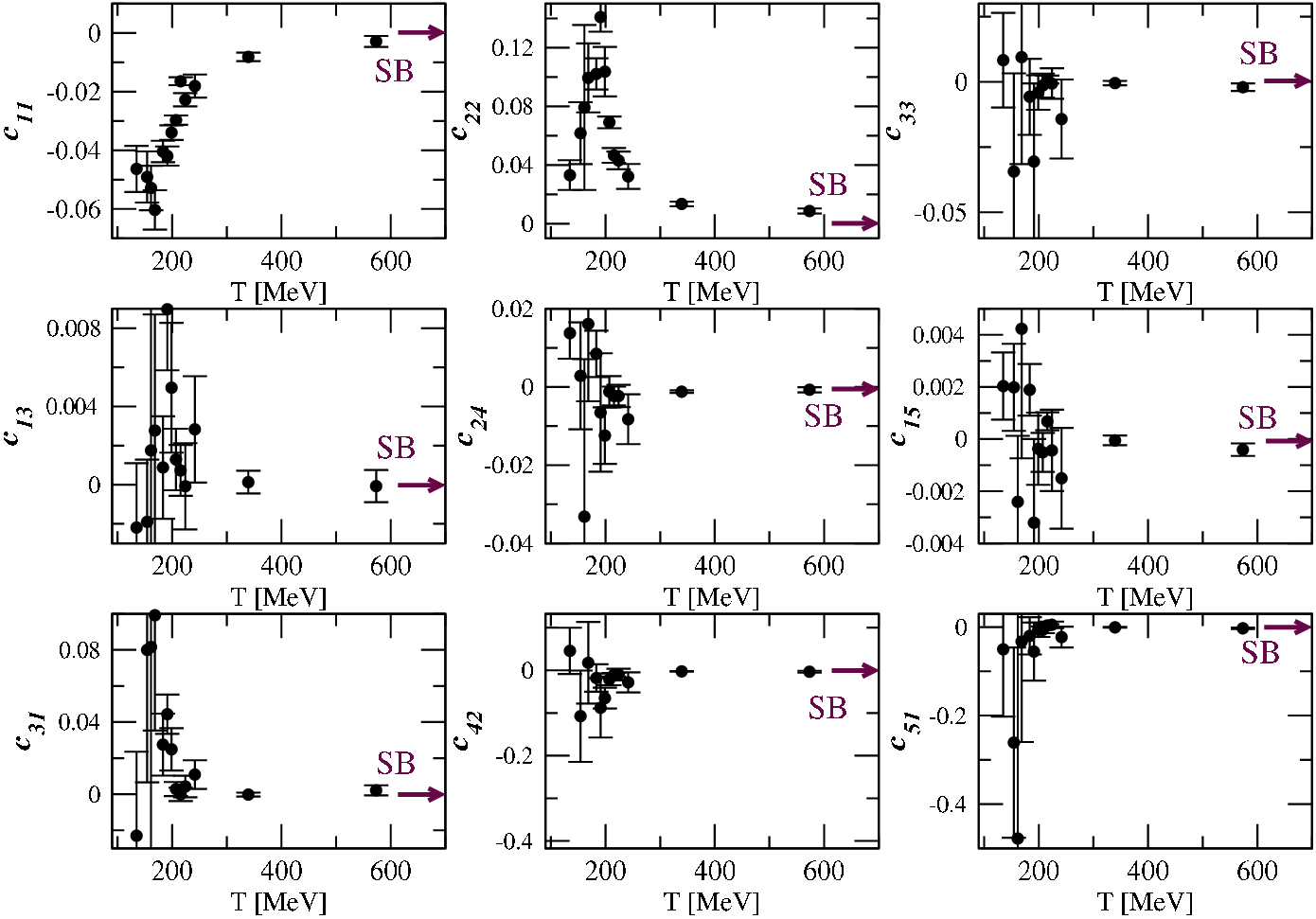}
\end{center}
\caption{Taylor expansion coefficients $c_{nm}(T)$ with $n,m\neq 0$ for $p/T^4$.}
\label{fig:c_m}
\end{figure}

Figure~\ref{fig:c_un} shows our results for the temperature dependence
of the $c_{n0}(T)$ and the $c_{0m}(T)$ coefficients. They all show rapid changes 
in the phase transition region and relatively quickly reach the 
Stefan-Boltzmann (SB) ideal gas values around $1.5T_c$ - $2T_c$.
Unsurprisingly, the errors of the higher order coefficients are larger
than the ones for the lowest order coefficients. 
They are worst for the sixth order coefficients $c_{60}(T)$ and $c_{06}(T)$. Although
the magnitude 
of the coefficients decreases 
with each order in the Taylor expansion, for $\bar{\mu}/T\sim 1$
the sixth order terms contribute a great deal of noise in the 
thermodynamic quantities at the present level of statistics.
Very similar conclusions can be made about the general behavior of
the rest of the pressure coefficients, $c_{nm}(T)$ with both $n,m\neq 0$, shown
in Fig.~\ref{fig:c_m}.
By comparison with the $c_{n0}(T)$ and $c_{0m}(T)$ coefficients, they are smaller
and so are their contributions to the various thermodynamic quantities.

\begin{figure}[h]
\epsfxsize=\textwidth
\begin{center}
\epsfbox{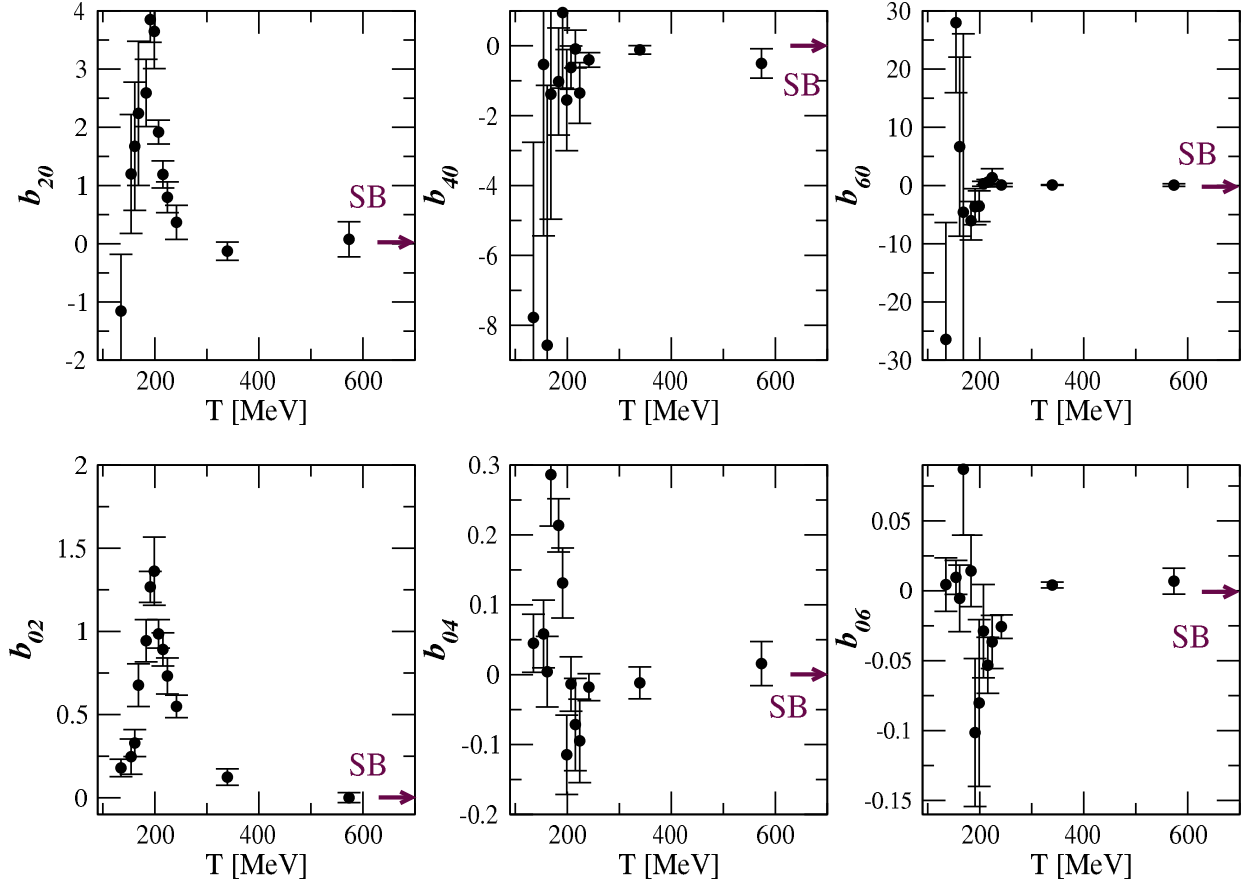}
\end{center}
\caption{Taylor expansion coefficients $b_{n0}(T)$ and $b_{0m}(T)$ for $I/T^4$.}
\label{fig:b_un}
\end{figure}

\begin{figure}[h]
\epsfxsize=\textwidth
\begin{center}
\epsfbox{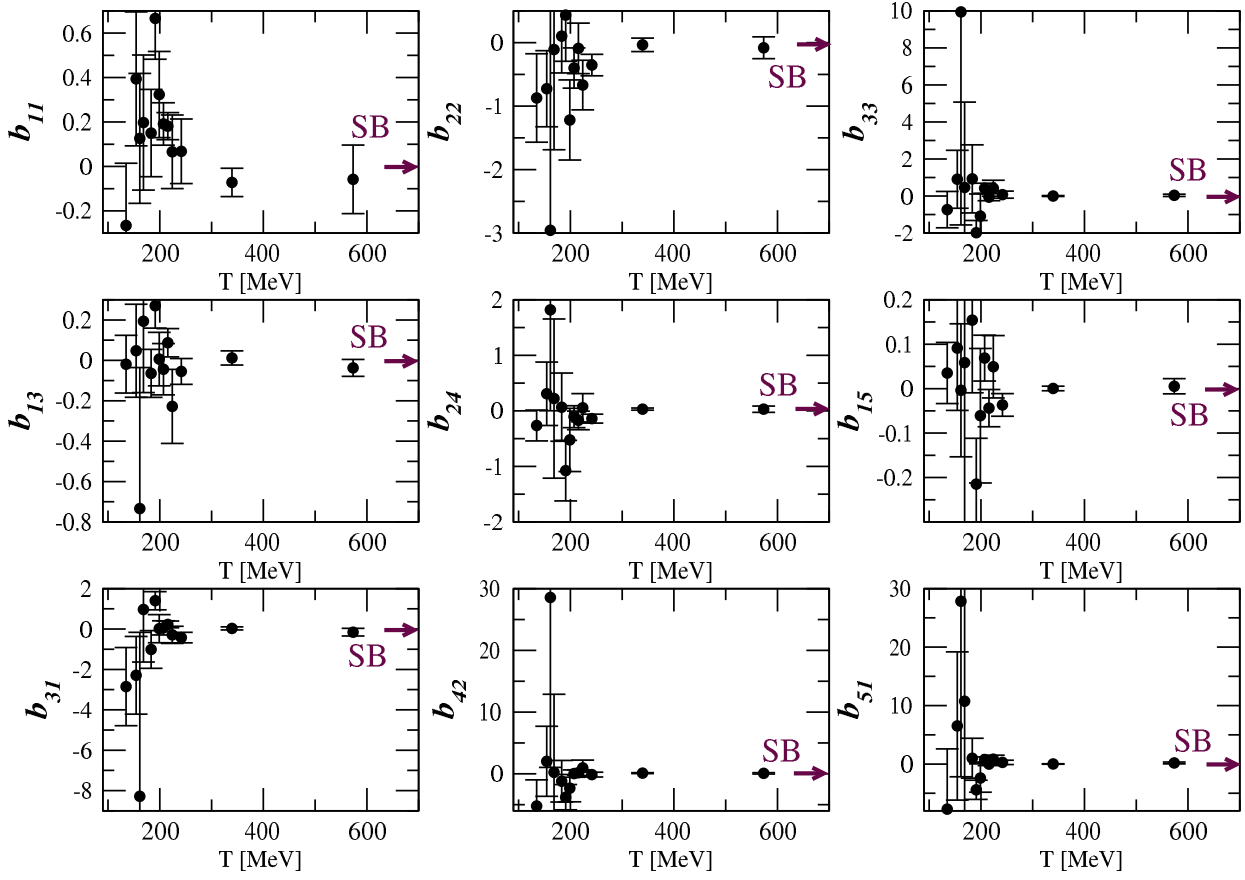}
\end{center}
\caption{Taylor expansion coefficients $b_{nm}(T)$ with $n,m\neq 0$ for $I/T^4$.}
\label{fig:b_m}
\end{figure}

\begin{figure}[h]
\begin{center}
\epsfxsize=84mm
\epsfbox{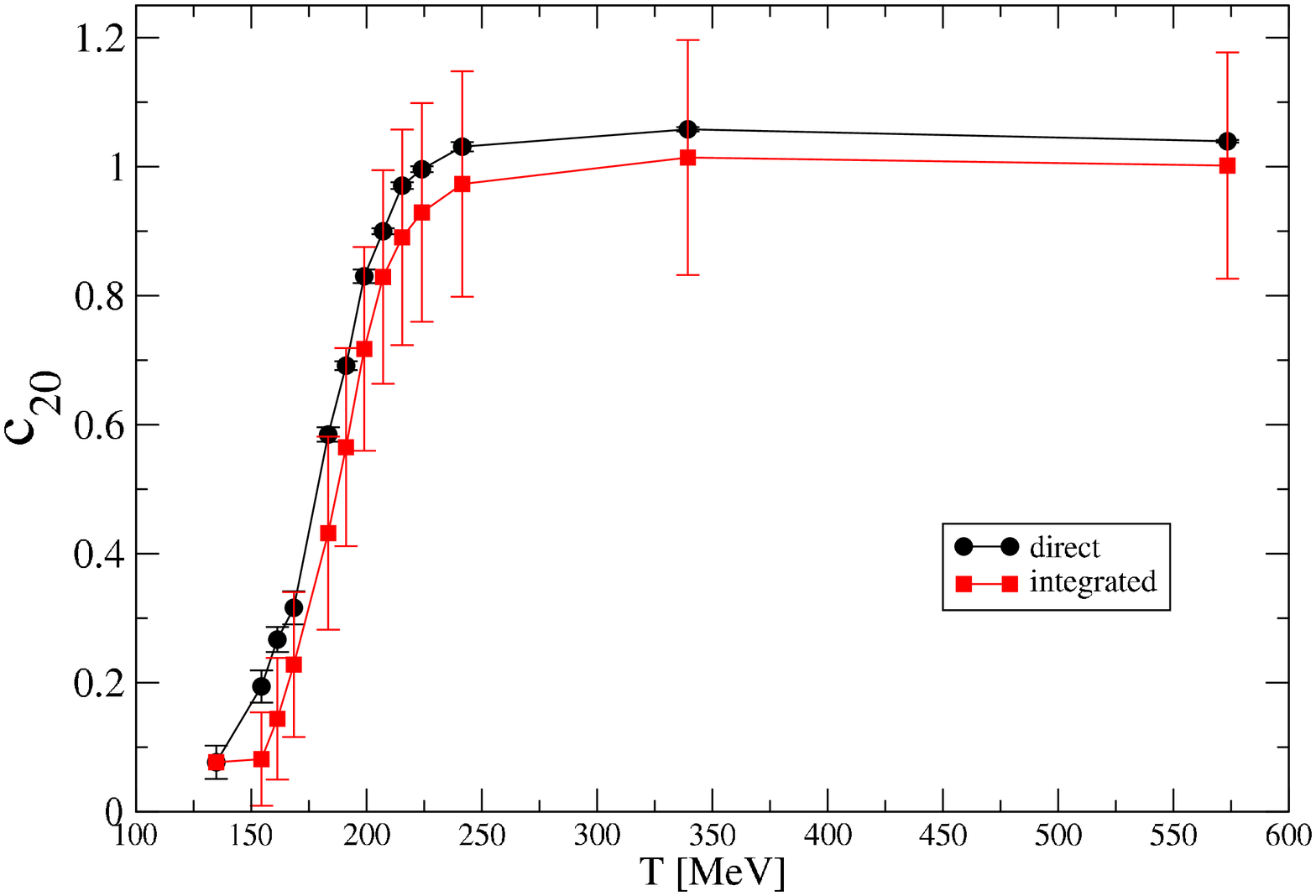}
\caption{Comparison between two different methods for calculating $c_{20}(T)$. The direct
method uses Eq.~(5) and the other method integrates $b_{20}(T)$ along the
trajectory of constant physics. The integral method produces significantly larger errors
than the direct one.}
\label{fig:comp}
\end{center}
\end{figure}

Figures~\ref{fig:b_un} and \ref{fig:b_m} show the coefficients in the
Taylor expansion of the interaction measure.
Here again we see the rapid changes/large fluctuations
around the transition region, the fast approach to the SB limit at high
temperatures and the increase in magnitude of the errors and 
the decrease in magnitude of the coefficients with each successive
order. In principle, each $c_{nm}(T)$ coefficient 
could be obtained from $b_{nm}(T)$
by integrating the latter along the trajectory of constant physics.
For example, 
in Fig.~\ref{fig:comp} the $c_{20}(T)$ coefficient obtained directly using
Eq.~(5) is compared to its value calculated by integrating $b_{20}(T)$.
The comparison shows that within the statistical 
errors the two results are the same. 
Similar
calculations were done for the rest of the coefficients and the consistency
between the results from the two methods
was satisfactory considering the large errors on the values 
obtained by integration.

\begin{figure}[ht]
\begin{tabular}{ll}
  \epsfxsize=80mm
  \epsfbox{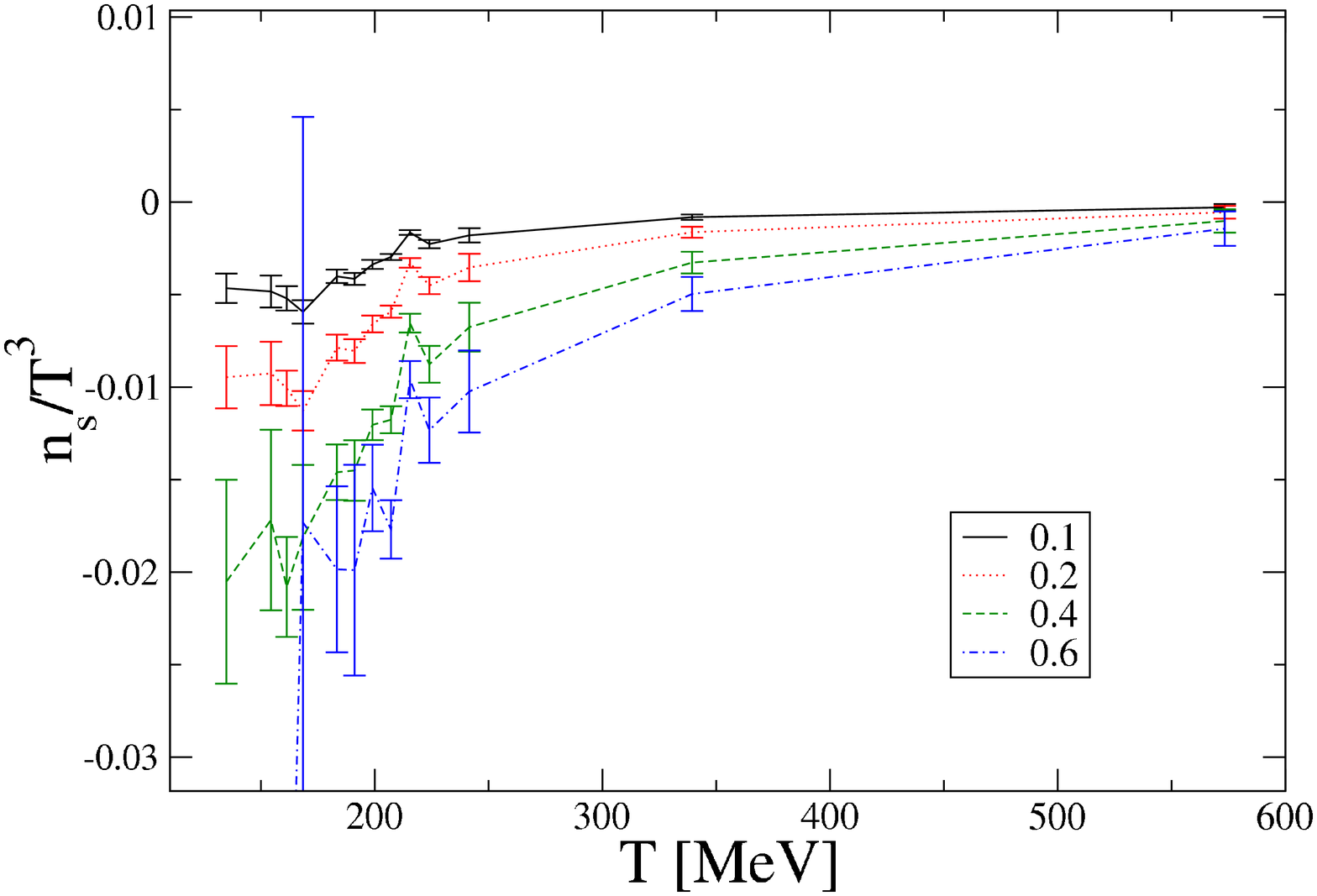}
  \epsfxsize=80mm
  \epsfbox{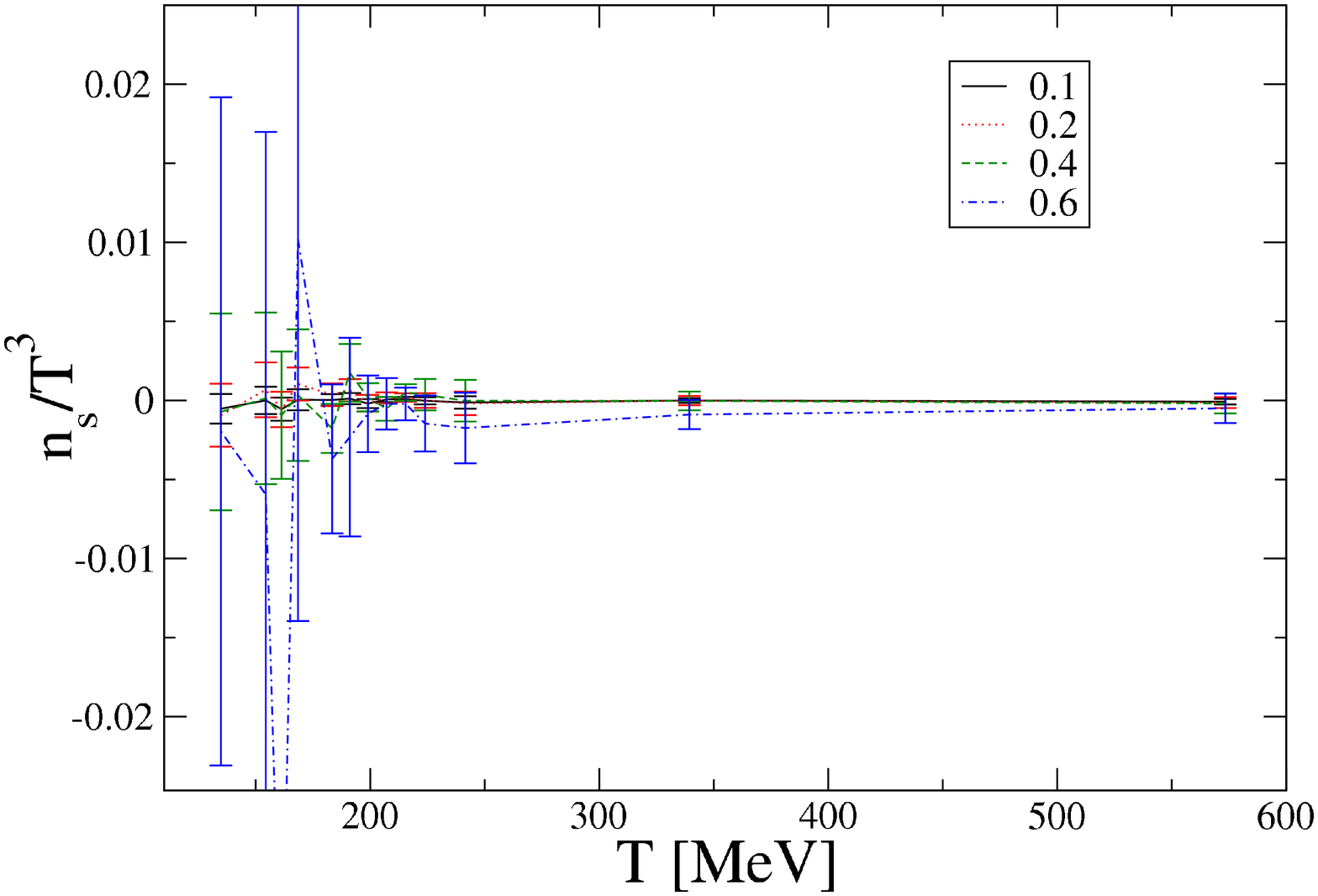}
\end{tabular}
\caption{The strange quark density $n_s/T^3$: left -- results with
$\bar{\mu}_h/T=0$; right -- tuned results. 
Different line styles
denote different values of $\bar{\mu}_l/T$.}
\label{fig:ns}
\end{figure}

Having determined the $c_{nm}(T)$ and $b_{nm}(T)$ coefficients we can now calculate 
the EOS to sixth order in the chemical potentials. 
We also determine the quark densities and various susceptibilities  
to fifth and fourth order, respectively.
Since we want to work at strange quark density $n_s=0$ to approximate the experimental
conditions, we tuned $\bar{\mu}_h/T$
along the trajectory of constant physics in order to achieve that condition within the
statistical error. Figure~\ref{fig:ns} (left)
shows, for several values of $\bar{\mu}_l/T$, that with $\bar{\mu}_h/T=0$ a
slightly negative $n_s$ is generated due to the nonzero $c_{n1}(T)$ terms.
After the introduction of an appropriate nonzero $\bar{\mu}_h/T$ 
for each studied temperature and $\bar{\mu}_l/T$, Fig.~\ref{fig:ns}
(right) shows our approximation of the condition $n_s=0$.
The effect of the tuning on thermodynamic quantities, 
other than $n_s/T^3$ itself, is small, because of the smallness 
of the ``mixed expansion
coefficients'' $c_{nm}(T)$ and $b_{nm}(T)$ for $n,m \neq 0$. For our level 
of statistics the typical effect is within the statistical errors
on the studied quantities.

\begin{figure}[ht]
\begin{tabular}{ll}
  \epsfxsize=80mm
  \epsfbox{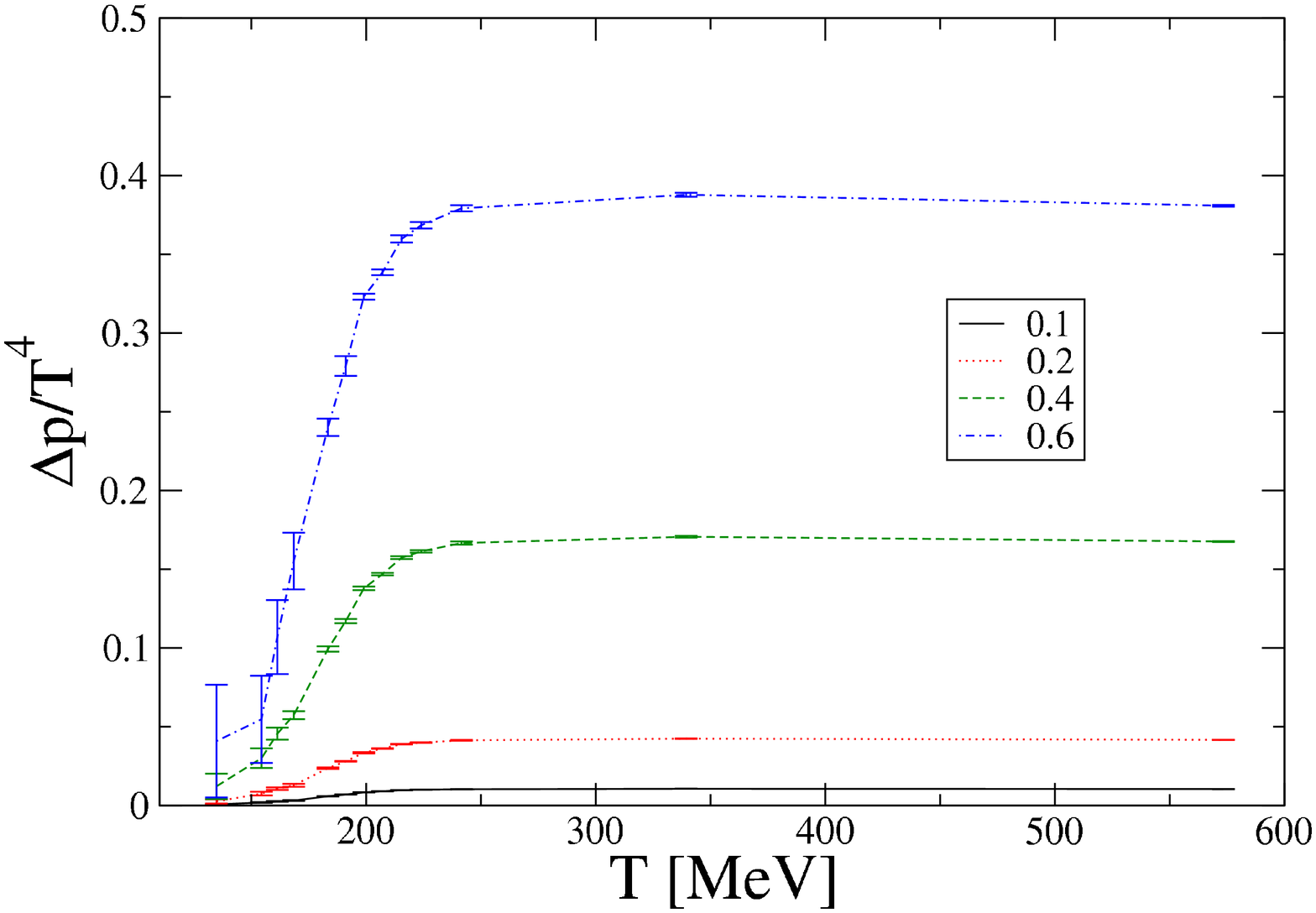}
  \epsfxsize=80mm
  \epsfbox{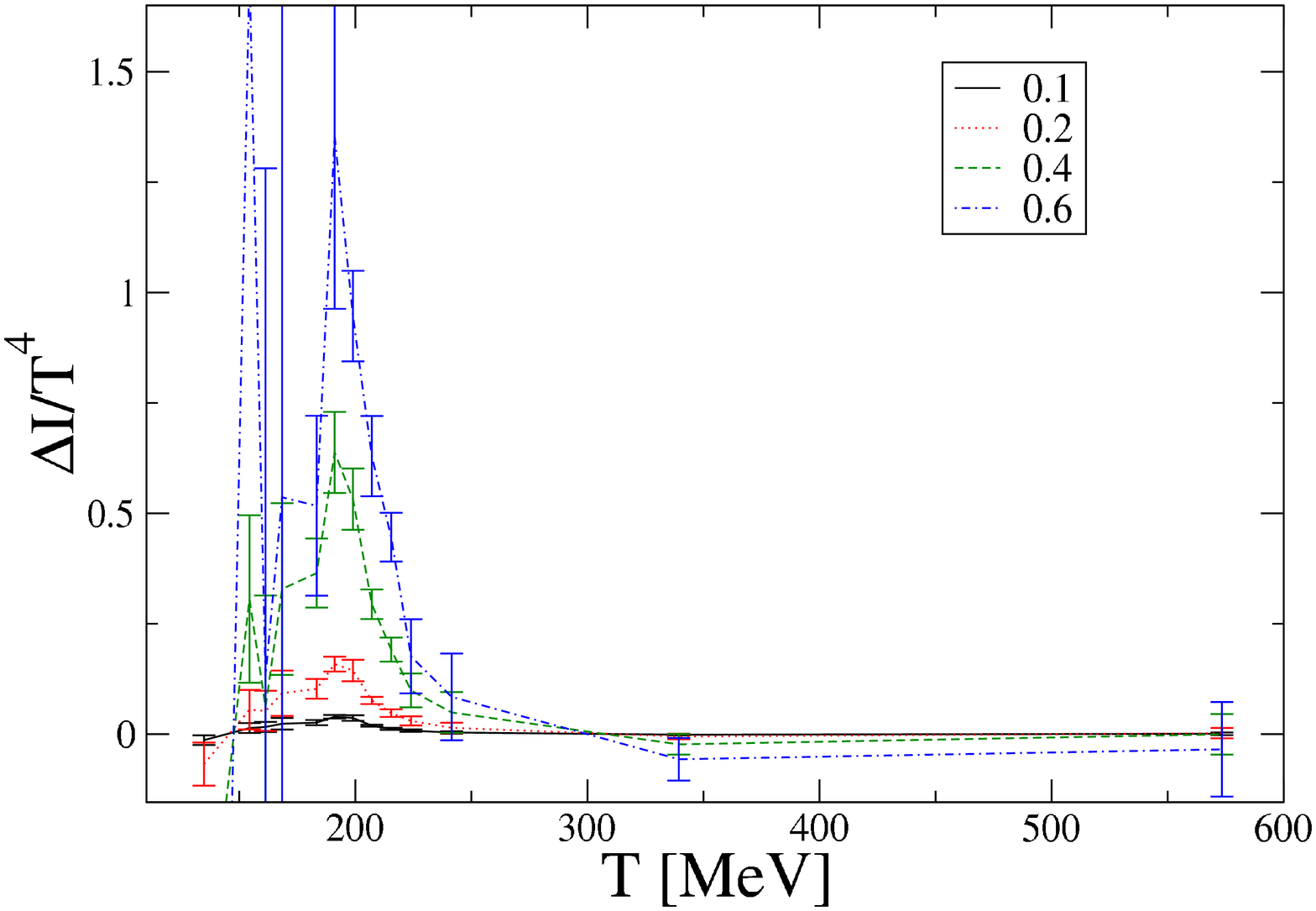}
\end{tabular}
\caption{Corrections to the pressure (left) and interaction measure (right)
at several values of $\bar{\mu}_l/T$. $\bar{\mu}_h/T$ is tuned such that
$n_s=0$ along the trajectory. }
\label{fig:dIdP}
\end{figure}

\begin{figure}[ht]
\epsfxsize=84mm
\epsfbox{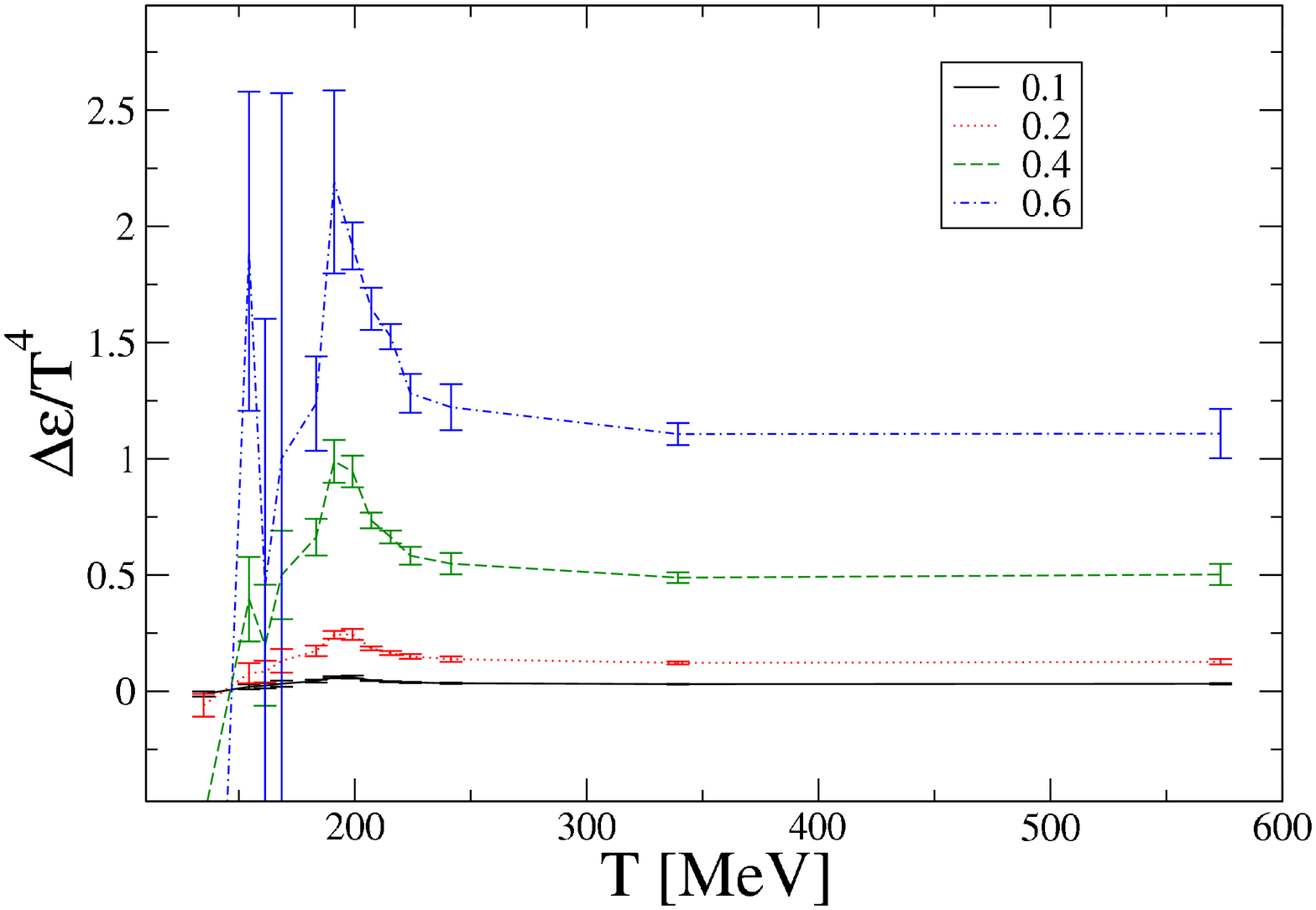}
\caption{Corrections to the energy density
at several values of $\bar{\mu}_l/T$. $\bar{\mu}_h/T$ is tuned such that
$n_s=0$ along the trajectory.}
\label{fig:dE}
\end{figure}

Figures~\ref{fig:dIdP} and \ref{fig:dE} show the corrections to the pressure,
interaction measure and energy density due to the presence of a nonzero 
$\bar{\mu}_l/T$. The correction to the pressure, for example, is the
difference 
$\Delta p/T^4 = p(\mu_{l,h}\neq0)/T^4 - p(\mu_{l,h}=0)/T^4$, which is Eq.~(5) minus the
zeroth order term $c_{00}(T)=p(\mu_{l,h}=0)/T^4$.
Similarly for the interaction measure and energy density, the corrections are
$
\Delta I/T^4 = I(\mu_{l,h}\neq0)/T^4 - I(\mu_{l,h}=0)/T^4$ and
$\Delta \varepsilon/T^4 = \varepsilon(\mu_{l,h}\neq0)/T^4 - \varepsilon(\mu_{l,h}=0)/T^4,
$  
which means again that the zeroth order terms are subtracted from 
the Taylor expansions for these quantities.
Qualitatively, our EOS results are similar to the previous two-flavor
studies \cite{Allton:2005gk}. The corrections to the thermodynamic quantities grow 
with increasing $\bar{\mu}_l/T$ and so do the statistical 
errors. The latter is due to the increasing contributions from higher
order terms, which are noisier than the lowest order terms.
\begin{figure}[ht]
\begin{tabular}{ll}
  \epsfxsize=80mm
  \epsfbox{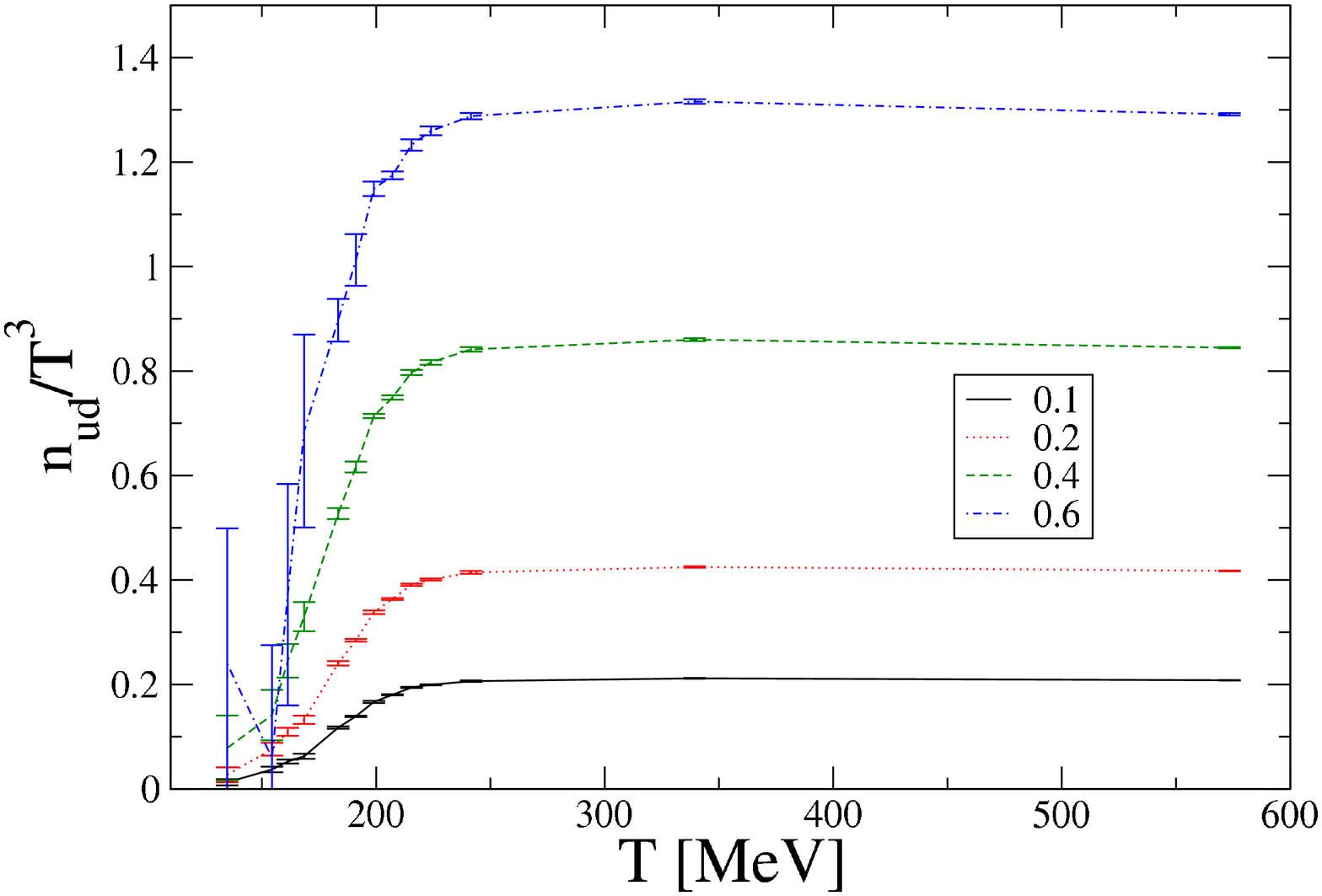}
  \epsfxsize=80mm
  \epsfbox{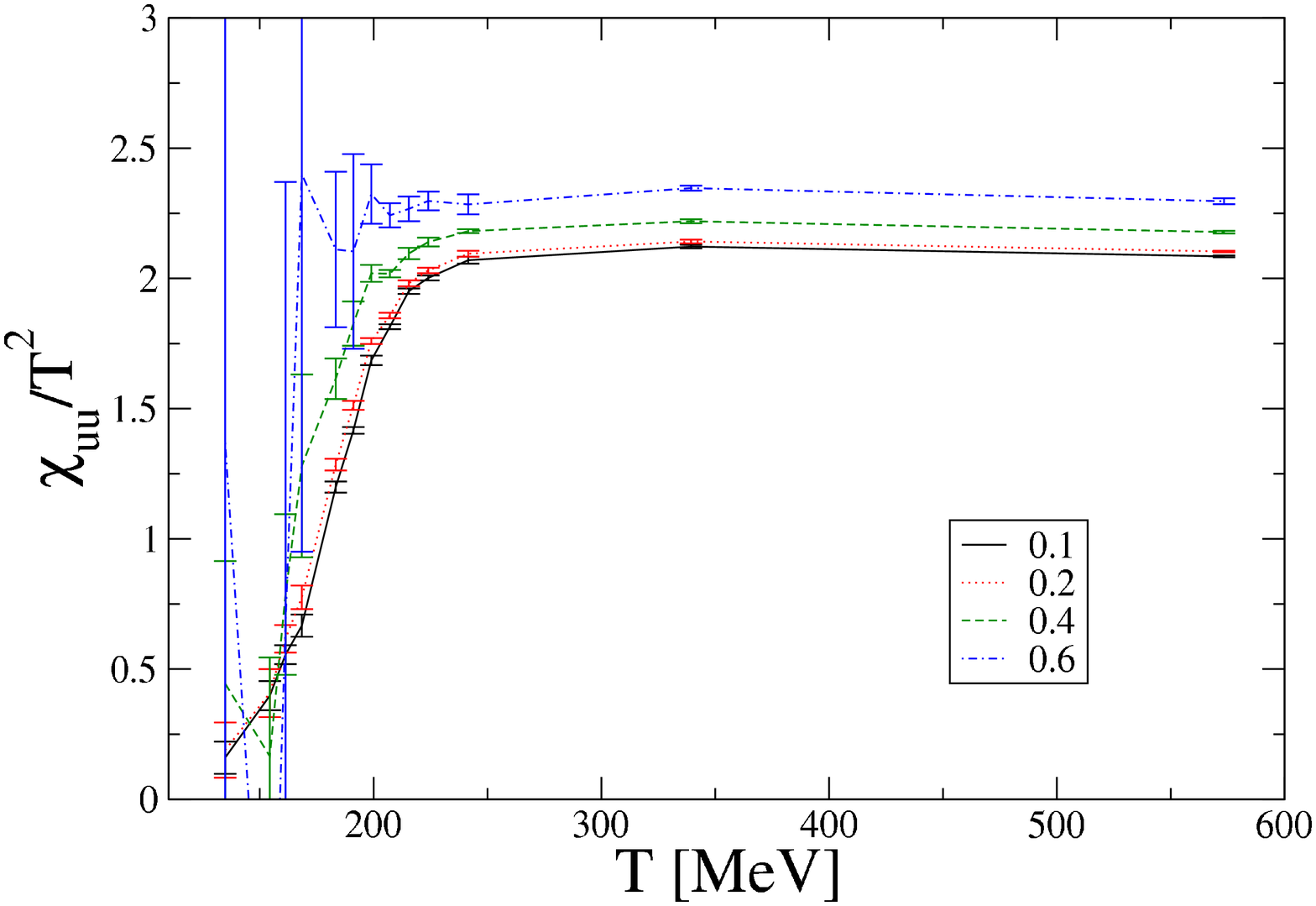}
\end{tabular}
\caption{Light quark density (left) and the light-light susceptibility (right)
at several values of $\bar{\mu}_l/T$. $\bar{\mu}_h/T$ is tuned such that
$n_s=0$ along the trajectory. }
\label{fig:nuchi}
\end{figure}
\begin{figure}[ht]
\begin{tabular}{ll}
  \epsfxsize=80mm
  \epsfbox{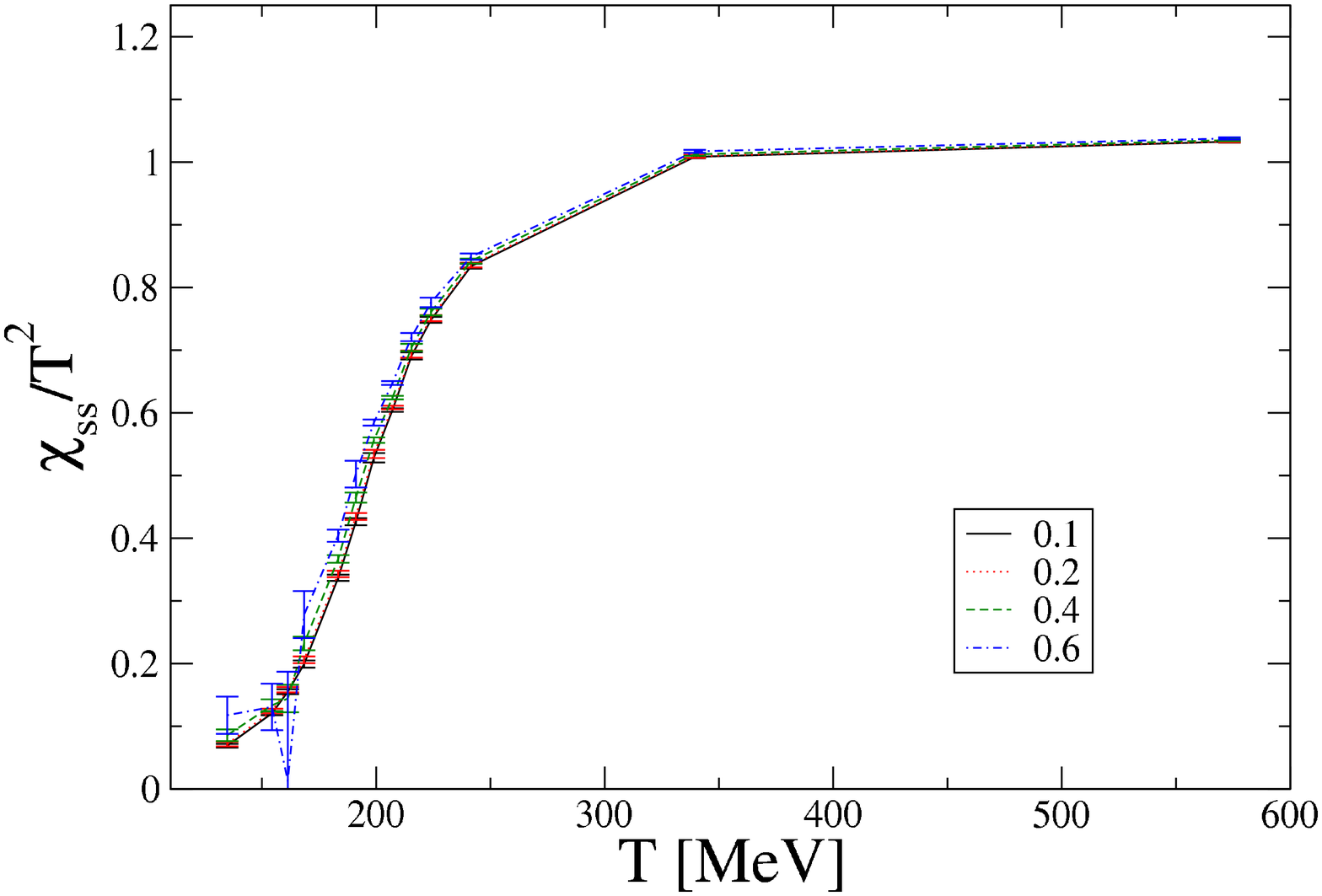}
  \epsfxsize=80mm
  \epsfbox{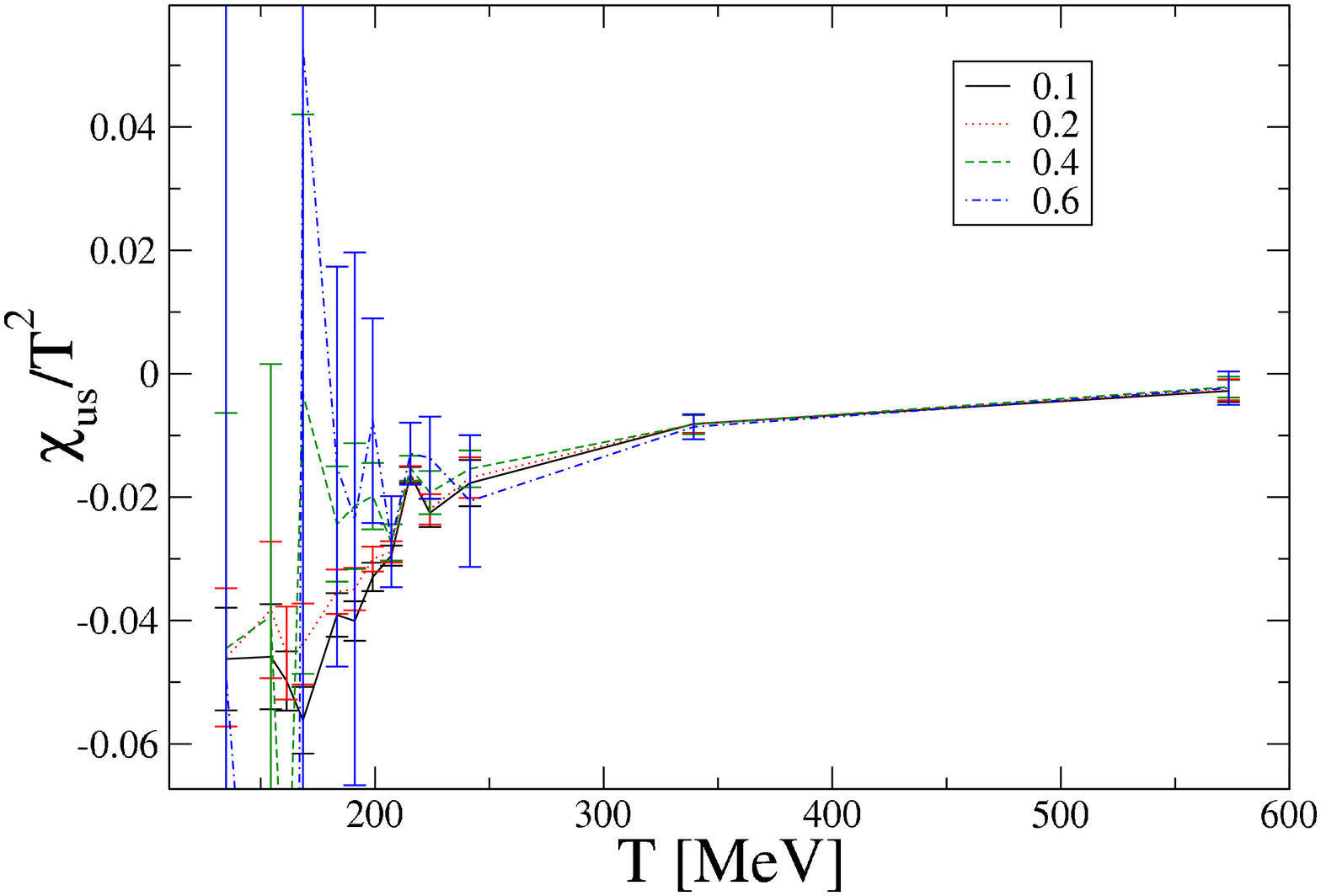}
\end{tabular}
\caption{Heavy-heavy (left) and heavy-light (right) susceptibilities
at several values of $\bar{\mu}_l/T$. $\bar{\mu}_h/T$ is tuned such that
$n_s=0$ along the trajectory. }
\label{fig:chichi}
\end{figure}
Figures~\ref{fig:nuchi} and \ref{fig:chichi} show that similar 
observations are true for the rest of the studied quantities:
the light quark density and the light-light, heavy-heavy and light-heavy
quark susceptibilities. Of these, the weakest dependence on $\bar{\mu}_l/T$ is shown
by the heavy-heavy susceptibility $\chi_{ss}$. A clear peak structure at 
the accessible $\bar{\mu}_l/T$ in the flavor diagonal light-light
quark susceptibility $\chi_{uu}$ would be a sign of reaching
the critical end point in the $\bar{\mu}-T$ plane. Our result does not
show  such a peak.
Considering the significant errors for larger values of $\bar{\mu}_l/T$,
it is difficult to say whether such a structure could be revealed
with higher statistics or if the critical $\bar{\mu}_l/T$ has not been 
reached here. In any case, reducing the statistical errors
and probably adding higher orders in the Taylor expansion 
would be the way to resolve that important problem.

\subsection{The isentropic EOS}
\begin{figure}[ht]
\begin{tabular}{ll}
  \epsfxsize=80mm
  \epsfbox{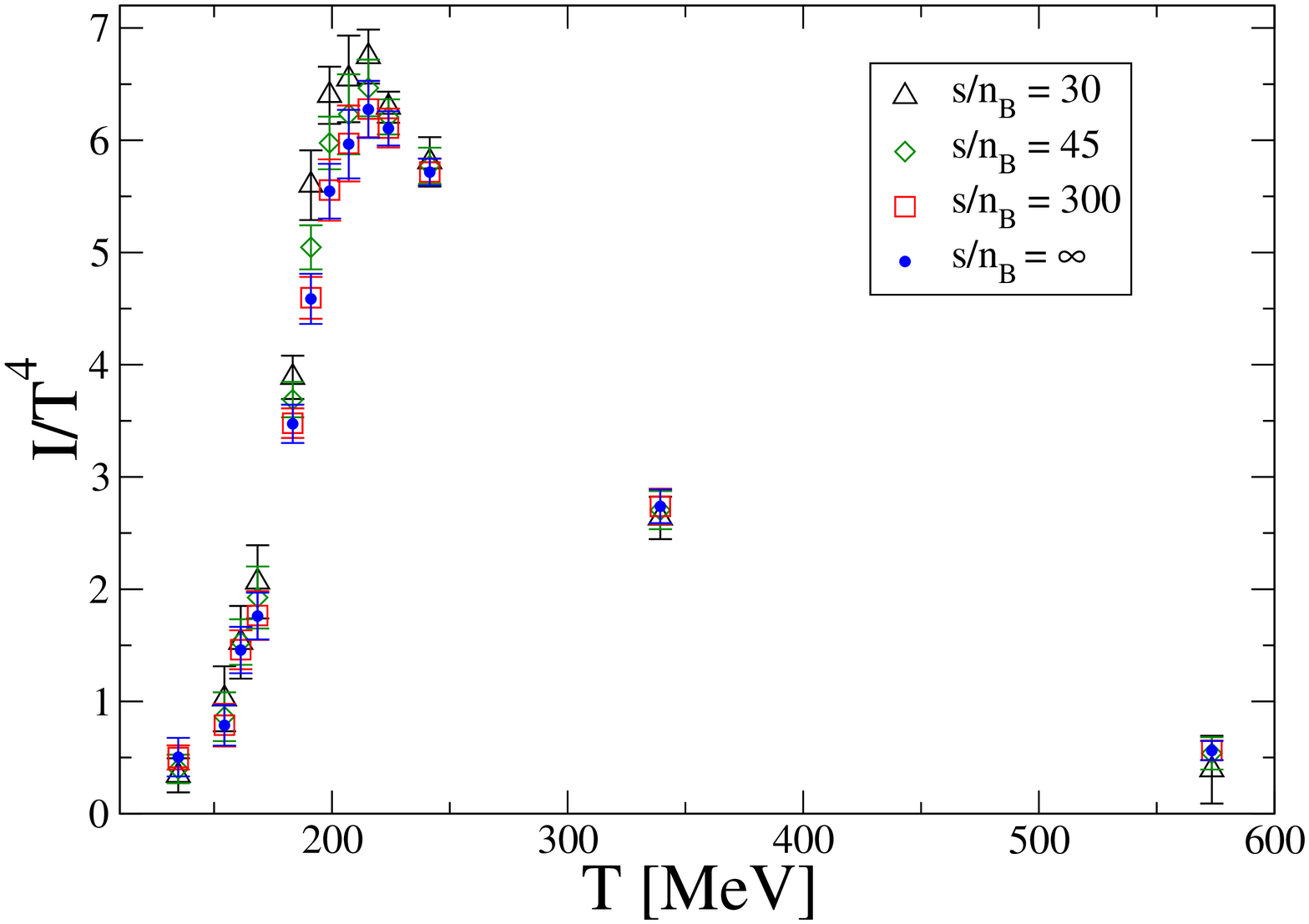}
  \epsfxsize=80mm
  \epsfbox{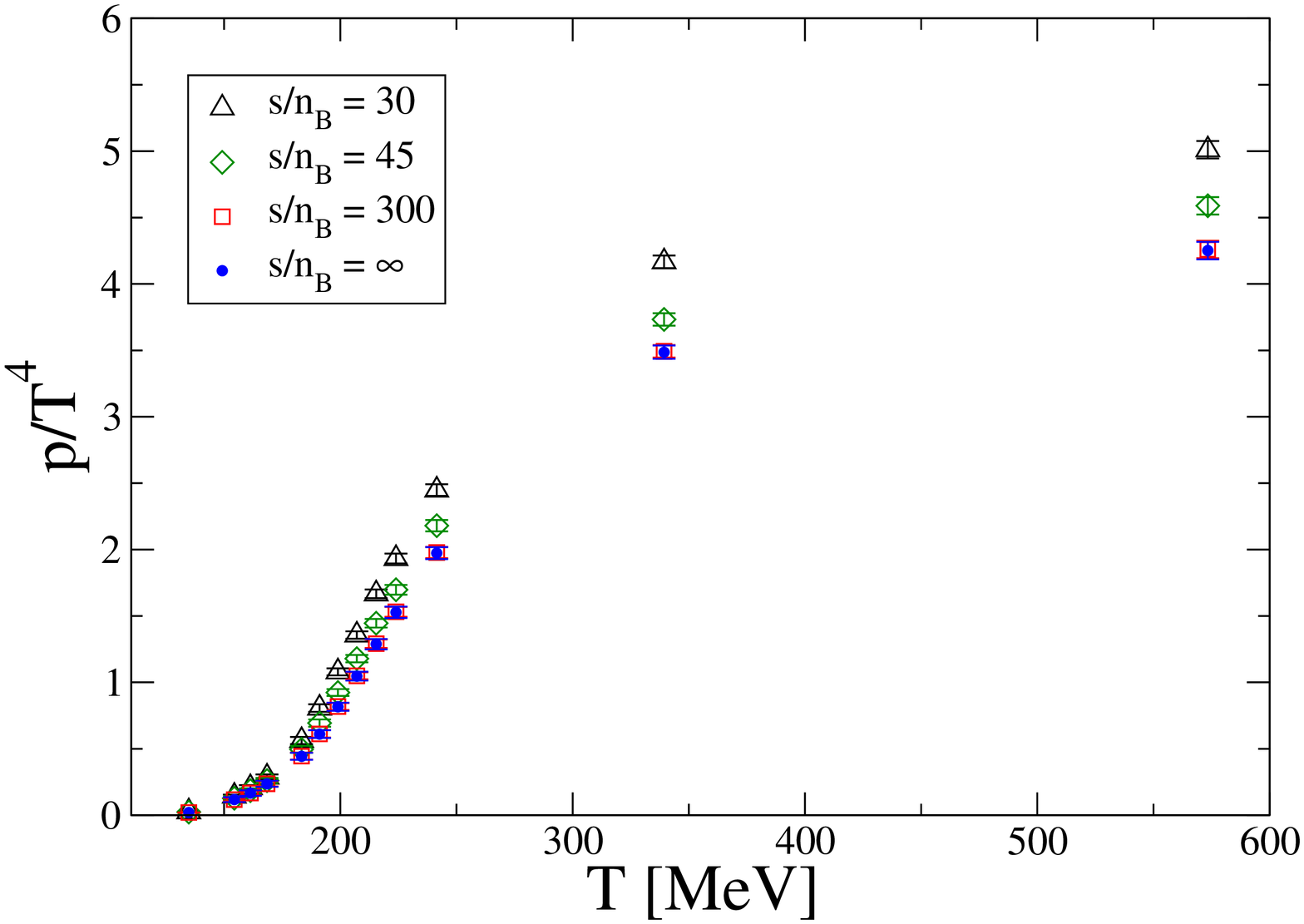}
\end{tabular}
\caption{Isentropic version of the interaction measure (left) and 
pressure (right) dependence on temperature at different finite values $s/b_B$
as described in the text. 
The case of zero chemical potential ($s/n_B=\infty$) is also shown.
These are the full results
for the quantities, not only the correction 
due to the nonzero chemical potential.}
\label{fig:IPisentr}
\end{figure}

\begin{figure}[ht]
\epsfxsize=83mm
\epsfbox{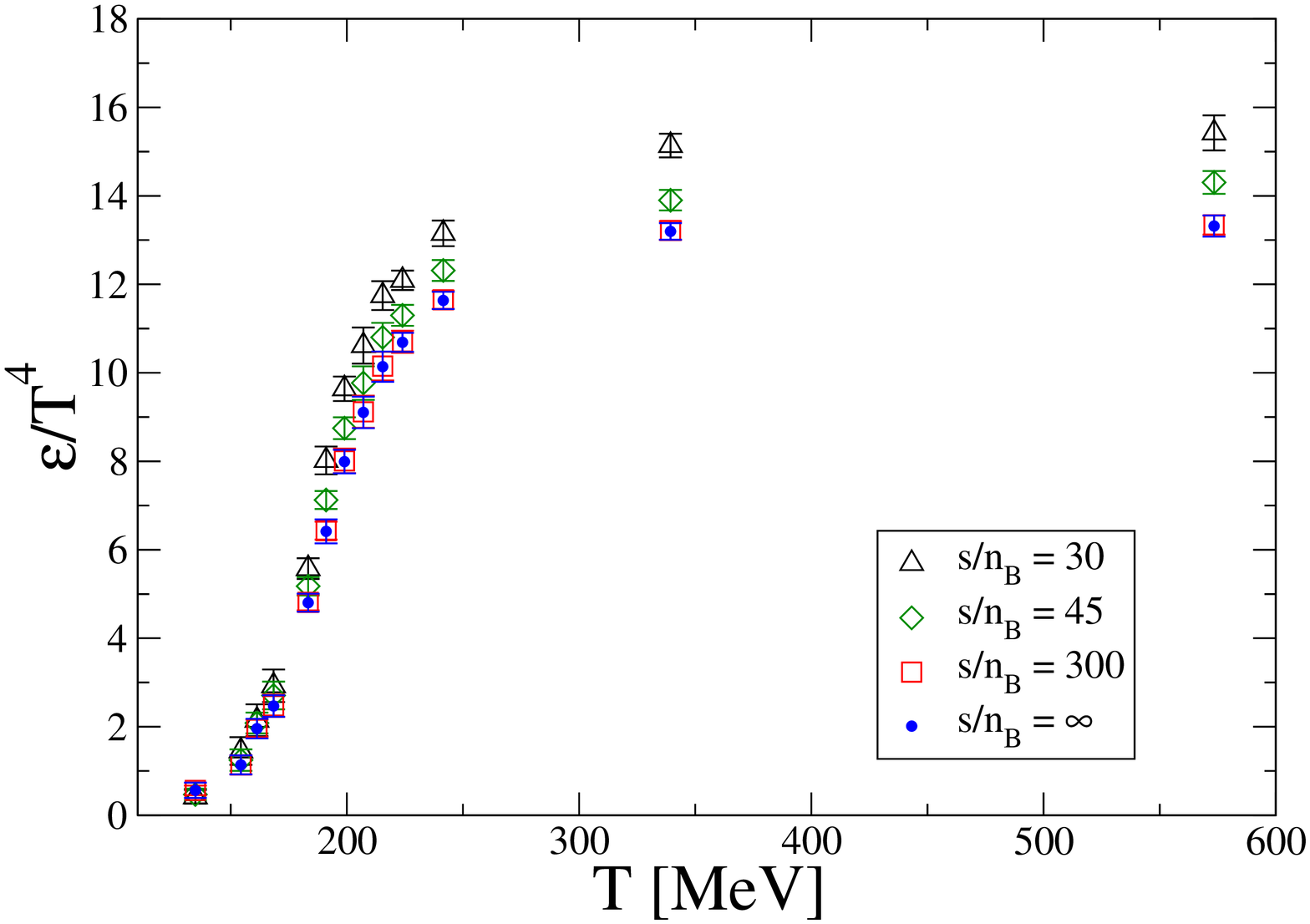}
\caption{Isentropic versions of the energy density 
dependence on temperature.}
\label{fig:Eisentr}
\end{figure}

\begin{figure}[ht]
\begin{tabular}{ll}
  \epsfxsize=80mm
  \epsfbox{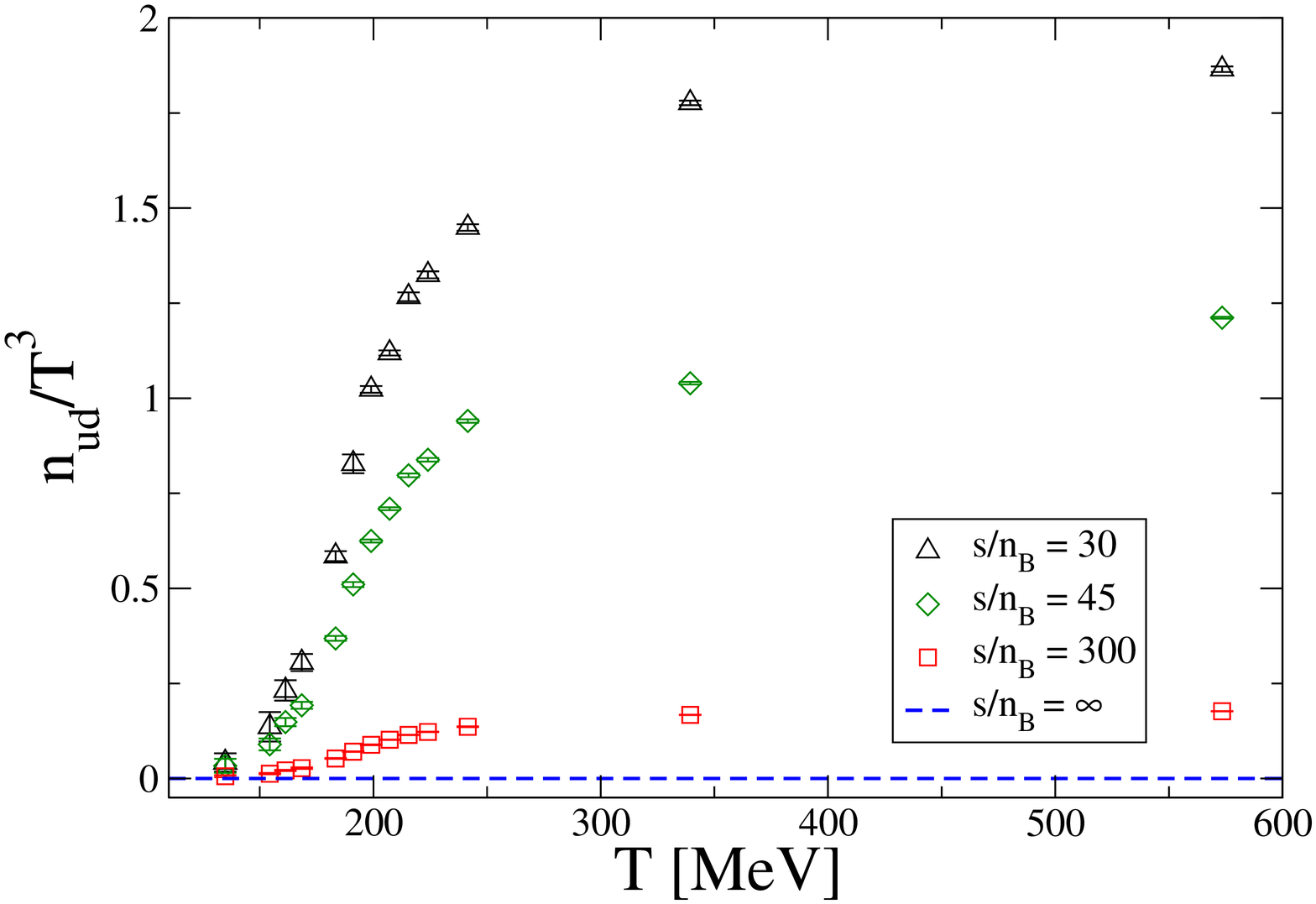}
  \epsfxsize=80mm
  \epsfbox{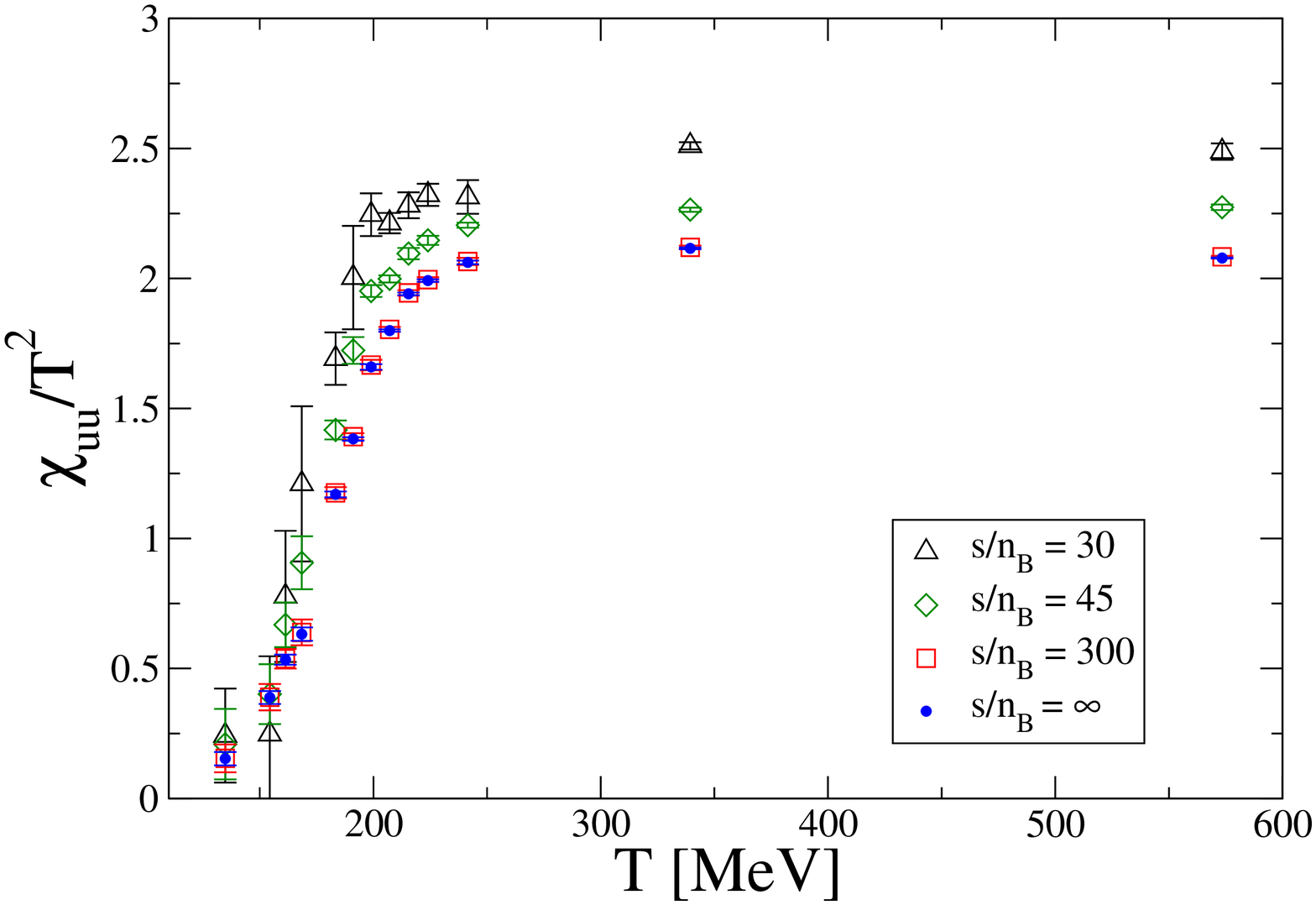}
\end{tabular}
\caption{Light quark density (left) and light-light susceptibility (right)
for different $s/n_B$.}
\label{fig:nchiisentr}
\end{figure}

\begin{figure}[h]
\begin{tabular}{ll}
  \epsfxsize=80mm
  \epsfbox{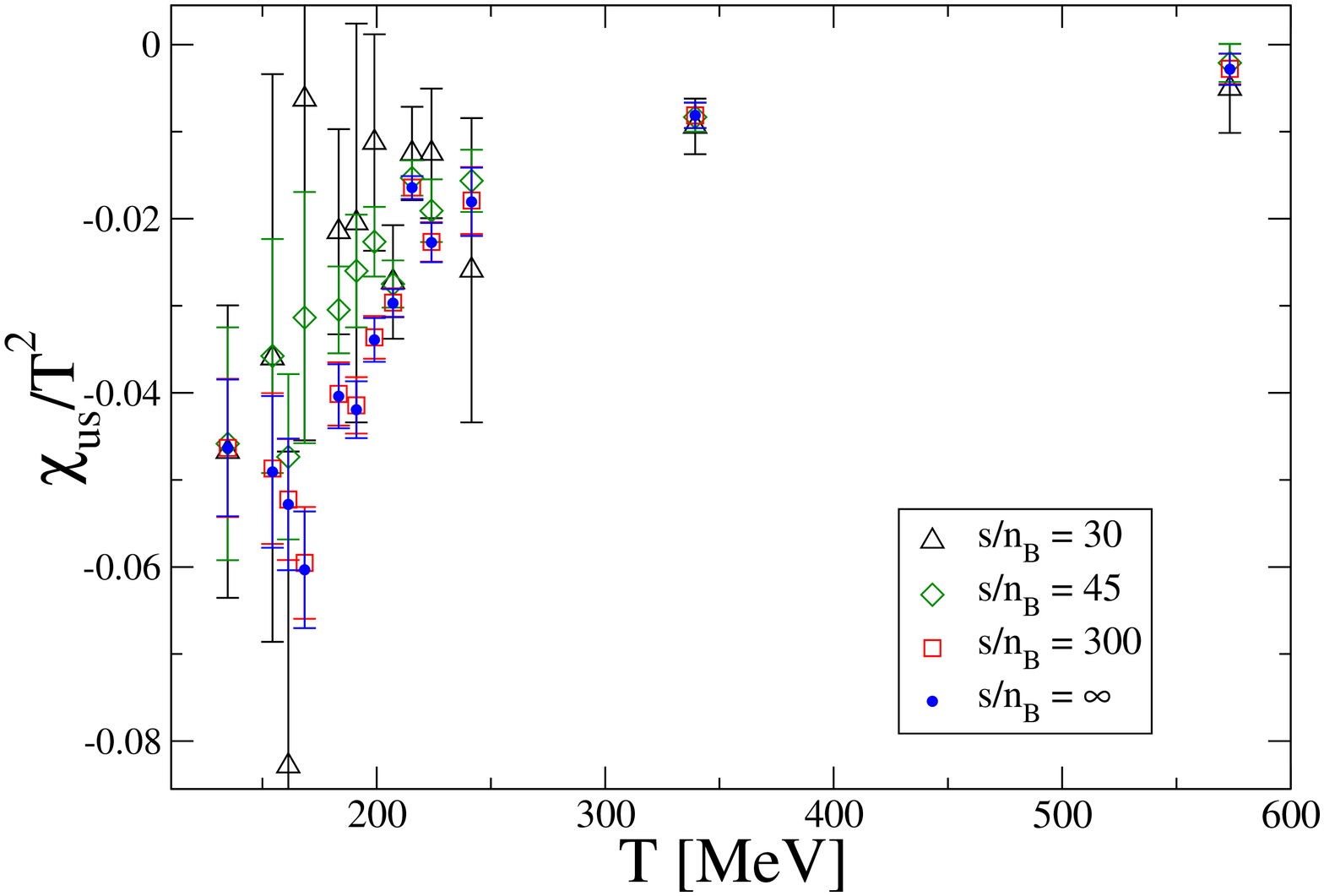}
  \epsfxsize=80mm
  \epsfbox{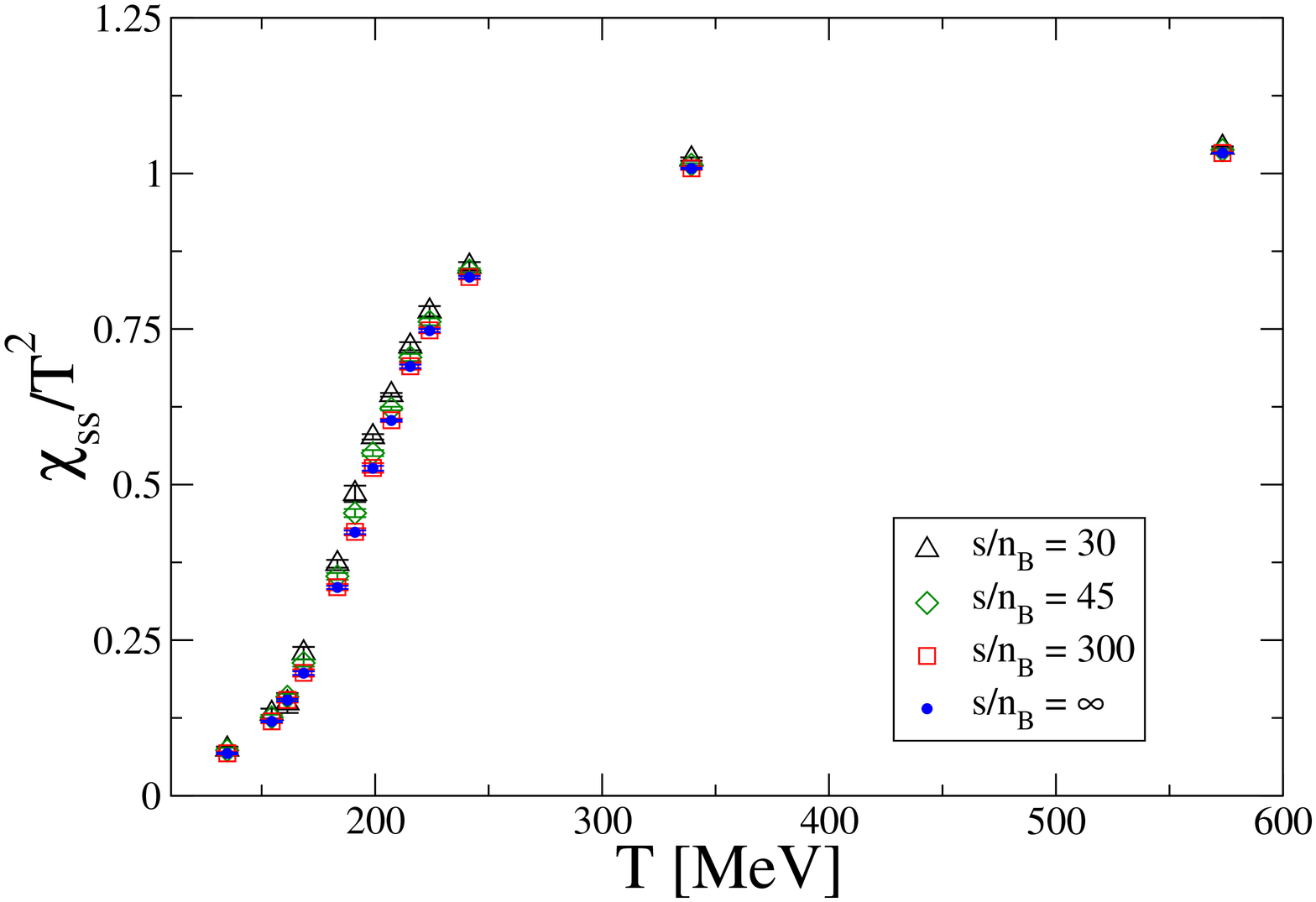}
\end{tabular}
\caption{Light-heavy (left) and heavy-heavy (right) susceptibilities
for different $s/n_B$.}
\label{fig:chichiisentr}
\end{figure}

The AGS, SPS and RHIC experiments produce matter 
which is expected to expand isentropically, {\it i.e.,} the entropy 
density $s$ and baryon number $n_B=n_{ud}/3$ 
both remain unchanged during the expansion. This implies that
$s/n_B$ remains constant. For the experiments mentioned, $s/n_B$ 
is approximately 30, 45 and 300 \cite{Ejiri:2005uv}, respectively.
In this subsection we present our results for the EOS and 
other thermodynamic quantities as calculated at nonzero chemical 
potential on  trajectories in the $\bar{\mu}-T$ space
with $s/n_B$ fixed at the values relevant to these experiments. 
\begin{figure}[ht]
\begin{tabular}{ll}
  \epsfxsize=83mm
  \epsfbox{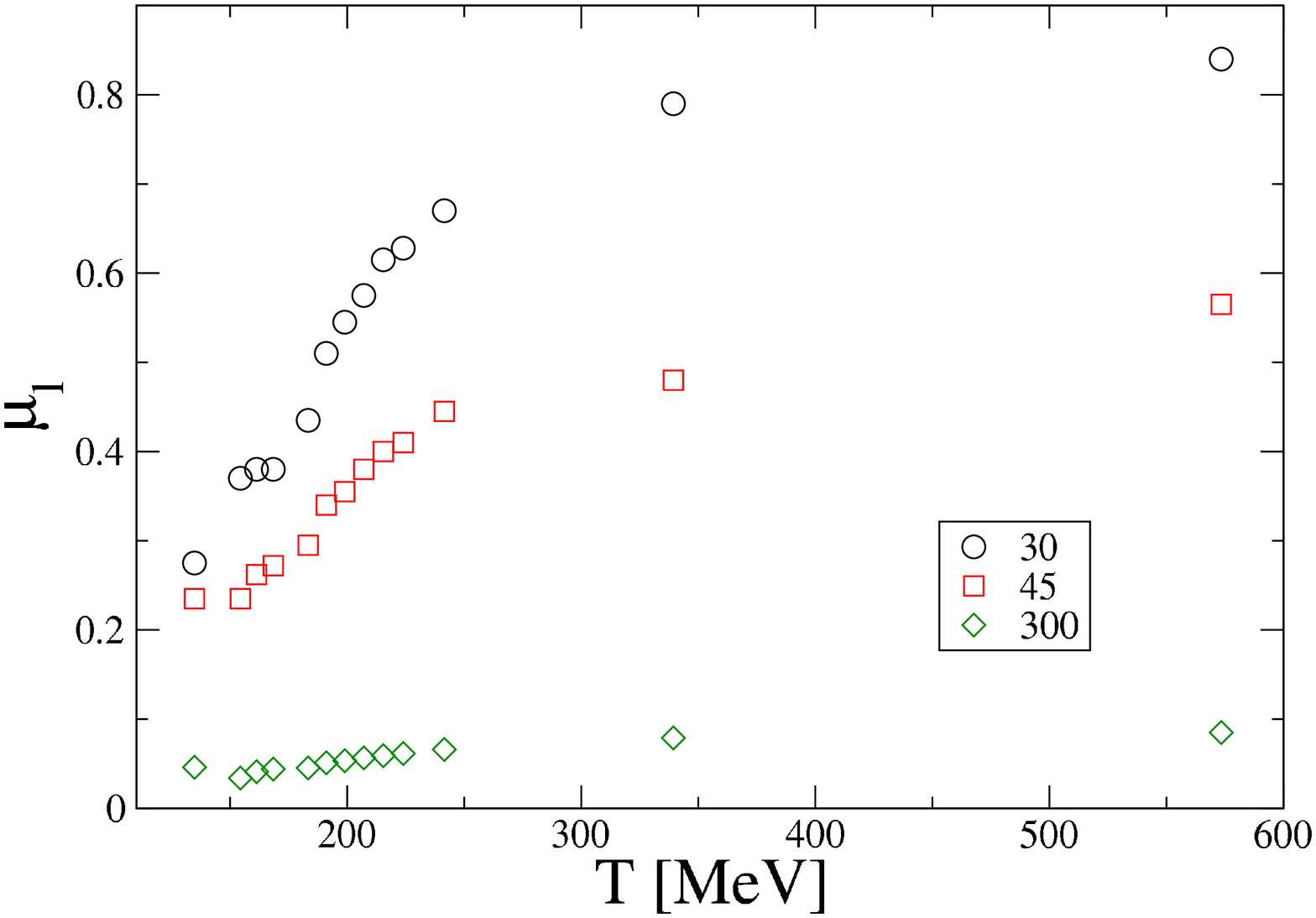}
  \epsfxsize=83mm
  \epsfbox{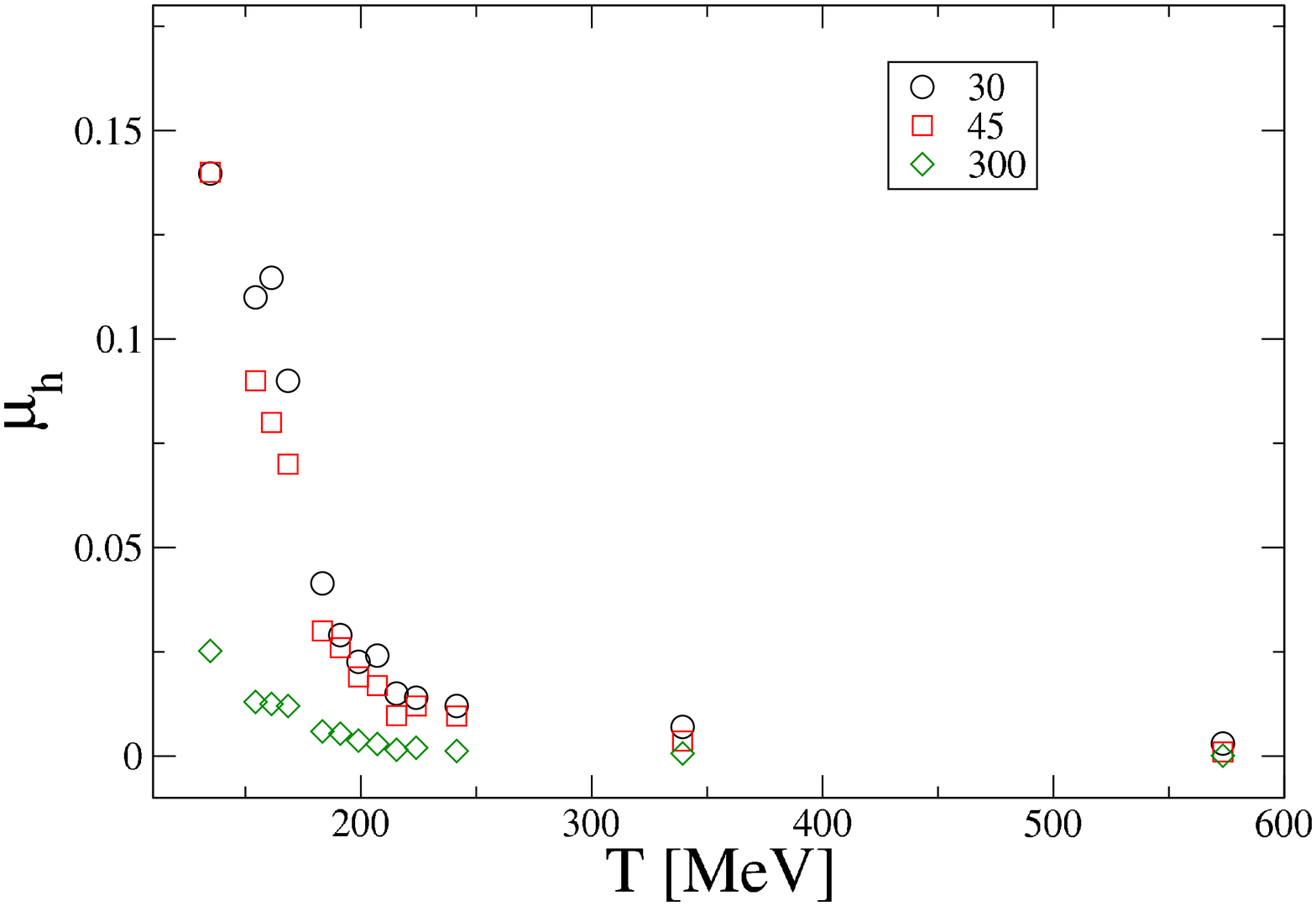}
\end{tabular}
\caption{The isentropic trajectories for different $s/n_B$.}
\label{fig:mtr}
\end{figure}
Figure~\ref{fig:mtr} shows the trajectories in the ($\mu_l$, $\mu_h$, $T$) 
space, obtained
by numerically solving the system
\bea
{s\over n_B}(\mu_l, \mu_h) &=& C\\
{n_s\over T^3} (\mu_l, \mu_h) &=& 0,
\eea
with $C=30$, 45, 300 for temperatures at which we have simulations. The tuning of the
parameters $\mu_l$ and $\mu_h$ is done until the deviations from 
$C$ and zero are no bigger than
the statistical errors of $s/ n_B$ and $n_s/T^3$, respectively.
After mapping the isentropic trajectories we use them to calculate the EOS,
the results for which are shown in Figs.~\ref{fig:IPisentr} 
and \ref{fig:Eisentr}. For comparison, we also
include the EOS result with $s/n_B = \infty$, which is the zero 
chemical potential case ($\mu_l=\mu_h=0$).
From the EOS results we conclude that in the studied range of $s/n_B$
the differences between the isentropic trajectories are not very large,
with the interaction measure least affected by the change in $s/n_B$.
Our results are again qualitatively very similar to the two-flavor isentropic
EOS study from \cite{Ejiri:2005uv}.
The isentropic results for $n_{ud}$, $\chi_{uu}$, $\chi_{us}$ and $\chi_{ss}$
are shown in Figs.~\ref{fig:nchiisentr} and \ref{fig:chichiisentr}.
It is interesting to note that $\chi_{uu}$ does not develop a peak 
structure on any of the isentropic trajectories. This means that all 
of the experiments work in the ranges of $s/n_B$ far from the critical 
end point, if such an end point exists at all for physical quark masses \cite{deForcrand:2006hh}. The light quark density
$n_{ud}$ looks most affected by the value of  $s/n_B$, and $\chi_{ss}$
is practically independent of it.

\section{Conclusions}
We have calculated the QCD equation of state for 2+1 flavors 
along a trajectory of constant physics and 
at nonzero chemical potential using the  
Taylor expansion method to sixth order in the chemical potential. 
The Taylor expansion coefficients
for the pressure and the interaction measure were determined directly
by measuring a set of fermionic and gluonic observables on the finite temperature 
ensembles along the trajectory. We used  Gaussian random
sources in the calculation of the 40 fermionic observables. The higher the order of the coefficients the noisier they
proved to be. Although the higher order coefficients have smaller magnitudes,
for increasing values of the chemical potential they contribute significantly 
to the statistical errors.
We tuned the heavy quark
chemical potential at each temperature studied in order to keep a 
vanishing strange quark density and have determined a number of
thermodynamic quantities at different values of the light quark chemical
potential for which the ratio 
$\bar{\mu}_l/T \lesssim 1$. Our corrections to the EOS
due to the nonzero chemical potential grow with the increasing values of
$\bar{\mu}_l/T$. However, not all thermodynamic quantities are equally affected
by the addition of a chemical potential.
Indeed, the heavy-heavy quark susceptibility is
practically independent of it.

We also have determined the isentropic versions of the EOS, the light 
quark densities and quark number susceptibilities, which are supposedly
most relevant for the current heavy-ion collision experiments. We found 
that the EOS is not strongly affected by changes in the ratio $s/n_B$,
which is in agreement with previous two-flavor results \cite{Ejiri:2005uv}.

\begin{center}
ACKNOWLEDGMENTS\\
\end{center}
This work was supported by the US Department of Energy under
  grants number DE-FG02-91ER-40628, DE-FC02-06ER-41446, DE-FG02-91ER-40661,
  DE-FC02-06ER-41443, DE-FC06-01ER-41437, DE-FG02-04ER-41298 and
  DE-FC02-06ER-41439 and by the US National Science Foundation under
  grants number  PHY05-555235, PHY04-56691, PHY05-55243, PHY05-55234,
  PHY04-56556, PHY05-55397 and PHY07-03296. Computations were performed at CHPC
(Utah), FNAL, FSU, IU, NCSA and UCSB.


\appendix
\section{Properties of the quark matrix derivatives}
We use the following identities for the fermion matrix
and its derivatives:
\begin{eqnarray}
M^{\dagger}(\mu) = \gamma_5 M(-\mu) \gamma_5,
\hspace{5mm} {\rm and} \hspace{5mm}
\frac{\partial^n M^{\dagger}}{\partial \mu^n}(\mu) =
(-1)^n \gamma_5 \frac{\partial^n M}{\partial \mu^n}(-\mu) \gamma_5.
\end{eqnarray}
Then, at $\mu=0$
\begin{eqnarray}
{\rm tr} \left( M^{-1} \frac{\partial^{n_1} M}{\partial \mu^{n_1}}
    M^{-1} \frac{\partial^{n_2} M}{\partial \mu^{n_2}} M^{-1}
    \cdots \right)^{*} = (-1)^{n_1+n_2+\cdots}
{\rm tr} \left( M^{-1} \frac{\partial^{n_1} M}{\partial \mu^{n_1}}
    M^{-1} \frac{\partial^{n_2} M}{\partial \mu^{n_2}} M^{-1}
    \cdots \right).
\end{eqnarray}
Because the terms in the $n$-th derivative satisfy
$n_1 + n_2 + \cdots = n$, we obtain
\begin{eqnarray}
\left( \frac{\partial^n \ln \det M}{\partial \mu^n} \right)^*
&=& (-1)^n \frac{\partial^n \ln \det M}{\partial \mu^n}; \\
\left( \frac{\partial^n {\rm tr} M^{-1}}{\partial \mu^n} \right)^*
&=& (-1)^n \frac{\partial^n {\rm tr} M^{-1}}{\partial \mu^n};\\
\left( \frac{\partial^n {\rm tr} M^{-1}\frac{dM}{du_0}}{\partial \mu^n}\right)^*
&=& (-1)^n \frac{\partial^n {\rm tr} M^{-1}\frac{dM}{du_0}}{\partial \mu^n};
\end{eqnarray}
i. e. all even derivatives are real and all odd ones are purely imaginary.
This means for example that
\be
{\rm Re}\left\langle L_2L_1L_1 \right\rangle = -\left\langle  {\rm Re}(L_2){\rm Im}(L_1){\rm Im}(L_1) \right\rangle,
\ee
and the real part of any observable containing odd number of 
odd derivatives is zero. 

Explicitly the derivatives of the asqtad fermion matrix are
\bea
\frac{\partial^n M}{\partial\mu^n} &=& \frac{1}{2} \eta_0(x) \left[ U_0^{(F)}(x) {\rm e}^\mu \delta_{x+\hat 0,y} -
 (-1)^nU_0^{(F)\dagger}(x-\hat{0}) {\rm e}^{-\mu} \delta_{x,y+\hat 0} \right. \\\nn 
 &&\left. + (3)^nU_0^{(L)}(x) {\rm e}^{3\mu} \delta_{x+3\hat 0,y} -
 (-3)^nU_0^{(L)\dagger}(x-3\hat{0}) {\rm e}^{-3\mu} \delta_{x,y+3\hat 0} \right]. 
\eea

\section{Algebraic techniques for the pressure}
The nonvanishing $c_{nm}(T)$ coefficients from second through sixth order are:
\begin{eqnarray}
c_{20} &\equiv& \frac{1}{2}\frac{\partial^2 (p/T^4)}{\partial (\mu_{l}/T)^2}
\biggr|_{\mu_l=0} =
\frac{1}{2} \frac{N_{t}}{N_{s}^3}\; {\cal A}_{20} \quad  \nonumber \\
c_{40} &\equiv& \frac{1}{4!} \frac{\partial^4 (p/T^4)}{\partial (\mu_{l}/T)^4}
\biggr|_{\mu_l=0} =
\frac{1}{4!} \frac{1}{N_{s}^3 N_{t}} ({\cal A}_{40} -3 {\cal A}_{20}^2) \quad 
\nonumber \\
c_{60} &\equiv& \frac{1}{6!} \frac{\partial^6 (p/T^4)}{\partial (\mu_{l}/T)^6}
\biggr|_{\mu_l=0} =
\frac{1}{6!} \frac{1}{N_{s}^3 N_{t}^3}
({\cal A}_{60} -15 {\cal A}_{40} {\cal A}_{20} +30 {\cal A}_{20}^3) \quad  \nonumber \\
\label{eq:dpmu0}
c_{02} &\equiv& \frac{1}{2}\frac{\partial^2 (p/T^4)}{\partial (\mu_{h}/T)^2}
\biggr|_{\mu_h=0} =
\frac{1}{2} \frac{N_{t}}{N_{s}^3}\; {\cal A}_{02} \quad  \nonumber \\
c_{04} &\equiv& \frac{1}{4!} \frac{\partial^4 (p/T^4)}{\partial (\mu_{h}/T)^4}
\biggr|_{\mu_h=0} =
\frac{1}{4!} \frac{1}{N_{s}^3 N_{t}} ({\cal A}_{04} -3 {\cal A}_{02}^2) \quad 
\nonumber \\
c_{06} &\equiv& \frac{1}{6!} \frac{\partial^6 (p/T^4)}{\partial (\mu_{h}/T)^6}
\biggr|_{\mu_h=0} =
\frac{1}{6!} \frac{1}{N_{s}^3 N_{t}^3}
({\cal A}_{06} -15 {\cal A}_{04} {\cal A}_{02} +30 {\cal A}_{02}^3) \quad \nonumber \\
\label{eq:dpmu0}
c_{11} &\equiv& \frac{1}{1!1!}\frac{\partial^2 (p/T^4)}{\partial (\mu_{l}/T)\partial (\mu_{h}/T) }
\biggr|_{\mu_{l,h}=0} =
\frac{N_{t}}{N_{s}^3}\; {\cal A}_{11} \quad  \nonumber \\
c_{31}&\equiv& \frac{1}{3!1!}\frac{\partial^4 (p/T^4)}{\partial^3 (\mu_{l}/T)\partial (\mu_{h}/T) }
\biggr|_{\mu_{l,h}=0} =
\frac{1}{3!1!}\frac{1}{N_{s}^3 N_{t}}\; ({\cal A}_{31} - 3\A_{20}\A_{11}) \quad  \nonumber \\
c_{13}&\equiv& \frac{1}{3!1!}\frac{\partial^4 (p/T^4)}{\partial (\mu_{l}/T)\partial^3 (\mu_{h}/T) }
\biggr|_{\mu_{l,h}=0} =
\frac{1}{3!1!}\frac{1}{N_{s}^3 N_{t}}\; ({\cal A}_{13} - 3\A_{02}\A_{11}) \quad  \nonumber \\
c_{22}&\equiv& \frac{1}{2!2!}\frac{\partial^4 (p/T^4)}{\partial^2 (\mu_{l}/T)\partial^2 (\mu_{h}/T) }
\biggr|_{\mu_{l,h}=0} =
\frac{1}{2!2!}\frac{1}{N_{s}^3 N_{t}}\; (\A_{22} - \A_{20}\A_{02} - 2\A_{11}^2) \quad  \nonumber \\
\vspace{1cm}
c_{42}&\equiv& \frac{1}{4!2!}\frac{\partial^6 (p/T^4)}{\partial^4 (\mu_{l}/T)\partial^2 (\mu_{h}/T) }
\biggr|_{\mu_{l,h}=0} =
\frac{1}{4!2!}\frac{1}{N_{s}^3 N_{t}^3}\; (\A_{42}-6\A_{20}\A_{22}-8\A_{11}\A_{31} -\A_{40}\A_{02}\nonumber \\
&&\hspace{5cm}+24\A_{20}\A_{11}^2+6\A_{02}\A_{20}^2)\quad  \nonumber \\
c_{24}&\equiv& \frac{1}{4!2!}\frac{\partial^6 (p/T^4)}{\partial^2 (\mu_{l}/T)\partial^4 (\mu_{h}/T) }
\biggr|_{\mu_{l,h}=0} =
\frac{1}{4!2!}\frac{1}{N_{s}^3 N_{t}^3}\; (\A_{24}-6\A_{02}\A_{22}-8\A_{11}\A_{13} -\A_{04}\A_{20}\nonumber \\
&&\hspace{5cm}+24\A_{02}\A_{11}^2+6\A_{20}\A_{02}^2)\quad  \nonumber \\
c_{51}&\equiv& \frac{1}{5!1!}\frac{\partial^6 (p/T^4)}{\partial^5 (\mu_{l}/T)\partial^1 (\mu_{h}/T) }
\biggr|_{\mu_{l,h}=0} =
\frac{1}{5!1!}\frac{1}{N_{s}^3 N_{t}^3}\; (\A_{51} -10\A_{31}\A_{20}-5\A_{40}\A_{11} +30\A_{11}\A_{20}^2)
\quad  \nonumber \\
c_{15}&\equiv& \frac{1}{5!1!}\frac{\partial^6 (p/T^4)}{\partial^1 (\mu_{l}/T)\partial^5 (\mu_{h}/T) }
\biggr|_{\mu_{l,h}=0} =
\frac{1}{5!1!}\frac{1}{N_{s}^3 N_{t}^3}\; (\A_{15} -10\A_{13}\A_{02}-5\A_{04}\A_{11} +30\A_{11}\A_{02}^2)
\quad  \nonumber \\
c_{33}&\equiv& \frac{1}{3!3!}\frac{\partial^6 (p/T^4)}{\partial^3 (\mu_{l}/T)\partial^3 (\mu_{h}/T) }
\biggr|_{\mu_{l,h}=0} =
\frac{1}{3!3!}\frac{1}{N_{s}^3 N_{t}^3}\; (\A_{33}- 3\A_{31}\A_{02} -3\A_{13}\A_{20} -9\A_{11}\A_{22}\nonumber \\
&&\hspace{5cm}+ 18\A_{20}\A_{02}\A_{11} + 12\A_{11}^3)\quad. \nonumber
\end{eqnarray}

To generate the above expressions for $c_{nm}$ we follow closely the technique 
given in \cite{Allton:2005gk}.
Let
\bea
\frac{\partial\,\ln{\cal Z}}{\partial \mu_l}&\equiv&\A_{10} = \langle L_1\rangle\\
\frac{\partial\,\ln{\cal Z}}{\partial \mu_h}&\equiv&\A_{01} = \langle H_1\rangle.
\eea
It can be shown that
\bea
\frac{\partial\A_{nm}}{\partial \mu_l}&=& \A_{n+1,m} - \A_{10}\A_{nm}\\
\frac{\partial\A_{nm}}{\partial \mu_h}&=& \A_{n,m+1} - \A_{01}\A_{nm},
\eea
where
\be
\A_{nm}\equiv\left\langle {\rm e}^{-L_{0}}{\rm e}^{-H_{0}}\frac{\partial^n {\rm e}^{L_{0}}}{\partial\mu_l^n}
\frac{\partial^m {\rm e}^{H_{0}}}{\partial\mu_h^m}\right\rangle.
\ee
Higher order derivatives of $\ln {\cal Z}$ at $\mu_{h,l}=0$ are zero if $n+m$ is odd,
 which can be shown to mean that, in this
case, $\A_{nm}=0$. An example for getting a higher order derivative using either Eq.~(B1) or Eq.~(B2):
\be
\frac{\partial^2\,\ln{\cal Z}}{\partial \mu_l\partial \mu_h}=\frac{\partial \A_{01}}{\partial \mu_l}=
\frac{\partial \A_{10}}{\partial \mu_h} =
\left.(\A_{11} - \A_{10}\A_{01})\right|_{\mu_{h,l}=0} = \A_{11}.
\ee 
Once an expression for $c_{nm}$ is obtained it is easy to get $c_{mn}$ by just interchanging 
$n$ and $m$ in the former.
The observables $\A_{nm}$ in terms of the operators
\be
L_n = \frac{n_l}{4} \frac{\partial^n \ln \det M_l}{\partial \mu_l^n}\hspace{1cm} {\rm and} \hspace{1cm}
H_m = \frac{n_h}{4} \frac{\partial^m \ln \det M_h}{\partial \mu_h^m},
\label{eq:basic}
\ee
 are
\begin{eqnarray}
{\cal A}_{20}  &=& 
\left\langle L_2 \right\rangle
+\left\langle L_1^2 \right\rangle
\nn\\
{\cal A}_{40}  &=& 
\left\langle L_4 \right\rangle
+4\left\langle L_3 L_1 \right\rangle
+3\left\langle L_2^2 \right\rangle
+6\left\langle L_2 L_1^2 \right\rangle
+\left\langle L_1^4 \right\rangle
\nn\\
{\cal A}_{60}  &=& 
\left\langle L_6 \right\rangle
+6\left\langle L_5 L_1 \right\rangle
+15\left\langle L_4 L_2 \right\rangle
+10\left\langle L_3^2 \right\rangle
+15\left\langle L_4 L_1^2 \right\rangle
+60\left\langle L_3 L_2 L_1 \right\rangle
\nonumber \\ && 
+15\left\langle L_2^3 \right\rangle
+20\left\langle L_3 L_1^3 \right\rangle
+45\left\langle L_2^2 L_1^2 \right\rangle
+15\left\langle L_2 L_1^4 \right\rangle
+\left\langle L_1^6 \right\rangle  \quad  \nn\\
{\cal A}_{02}  &=& 
\left\langle H_2 \right\rangle
+\left\langle H_1^2 \right\rangle
\nn\\
{\cal A}_{04}  &=& 
\left\langle H_4 \right\rangle
+4\left\langle H_3 H_1 \right\rangle
+3\left\langle H_2^2 \right\rangle
+6\left\langle H_2 H_1^2 \right\rangle
+\left\langle H_1^4 \right\rangle
\nn\\
{\cal A}_{06}  &=& 
\left\langle H_6 \right\rangle
+6\left\langle H_5 H_1 \right\rangle
+15\left\langle H_4 H_2 \right\rangle
+10\left\langle H_3^2 \right\rangle
+15\left\langle H_4 H_1^2 \right\rangle
+60\left\langle H_3 H_2 H_1 \right\rangle
\nonumber \\ && 
+15\left\langle H_2^3 \right\rangle
+20\left\langle H_3 H_1^3 \right\rangle
+45\left\langle H_2^2 H_1^2 \right\rangle
+15\left\langle H_2 H_1^4 \right\rangle
+\left\langle H_1^6 \right\rangle  \quad \nn\\
\A_{11} &=&\left\langle L_1H_1 \right\rangle \nn\\
{\cal A}_{22}& =& \left\langle L_2 H_2\right\rangle 
 +\left\langle L_2 H_1^2\right\rangle 
 + \left\langle L_1^2 H_2 \right\rangle 
+ \left\langle L_1^2 H_1^2 \right\rangle \nn\\
\A_{31} &=& \left\langle L_3H_1 \right\rangle  
+3\left\langle L_2L_1H_1  \right\rangle
+\left\langle L_1^3H_1 \right\rangle  \nn\\
\A_{13} &=& \left\langle H_3L_1 \right\rangle  
+3\left\langle H_2H_1L_1  \right\rangle
+\left\langle H_1^3L_1 \right\rangle  \nn\\
\A_{42}&=& 
\left\langle L_4H_2 \right\rangle
+4\left\langle L_3 L_1H_2 \right\rangle
+3\left\langle L_2^2H_2 \right\rangle
+6\left\langle L_2 L_1^2H_2 \right\rangle
+\left\langle L_1^4H_2 \right\rangle
\nn\\
&&
+\left\langle L_4H_1^2 \right\rangle
+4\left\langle L_3 L_1H_1^2 \right\rangle
+3\left\langle L_2^2H_1^2 \right\rangle
+6\left\langle L_2 L_1^2H_1^2 \right\rangle
+\left\langle L_1^4H_1^2 \right\rangle
\nn\\
\A_{24}&=& 
\left\langle H_4L_2 \right\rangle
+4\left\langle H_3 H_1L_2 \right\rangle
+3\left\langle H_2^2L_2 \right\rangle
+6\left\langle H_2 H_1^2L_2 \right\rangle
+\left\langle H_1^4L_2 \right\rangle
\nn\\
&&
+\left\langle H_4L_1^2 \right\rangle
+4\left\langle H_3 H_1L_1^2 \right\rangle
+3\left\langle H_2^2L_1^2 \right\rangle
+6\left\langle H_2 H_1^2L_1^2 \right\rangle
+\left\langle H_1^4L_1^2 \right\rangle
\nn\\
\A_{51}&=&
\left\langle L_5H_1 \right\rangle+
5\left\langle L_4L_1H_1 \right\rangle+
10\left\langle L_3L_2H_1 \right\rangle+
10\left\langle L_3L_1^2H_1 \right\rangle+
15\left\langle L_2^2L_1H_1 \right\rangle
\nn\\&&+
10\left\langle L_2L_1^3H_1 \right\rangle+
\left\langle L_1^5H_1 \right\rangle\nn\\
\A_{15}&=&
\left\langle H_5L_1 \right\rangle+
5\left\langle H_4H_1L_1 \right\rangle+
10\left\langle H_3H_2L_1 \right\rangle+
10\left\langle H_3H_1^2L_1 \right\rangle+
15\left\langle H_2^2H_1L_1 \right\rangle
\nn\\&&+
10\left\langle H_2H_1^3L_1 \right\rangle+
\left\langle H_1^5L_1 \right\rangle \nn\\
\A_{33}&=&\left\langle
(L_3+3L_2L_1 +L_1^3)
(H_3+3H_2H_1 +H_1^3)\right\rangle. \nn
\end{eqnarray}
The observables $L_n$ and $H_m$ include the quark matrix 
$M(=M_{l,h})$ derivatives with respect to $\mu (=\mu_{l,h})$, which have the following 
form:
\begin{small}
\begin{eqnarray}
\frac{\partial \ln \det M}{\partial \mu}
 &=& 
{\rm tr} \left( M^{-1} \frac{\partial M}{\partial \mu} \right) 
\label{eq:dermu1} \\
\frac{\partial^2 \ln \det M}{\partial \mu^2}
 &=& 
{\rm tr} \left( M^{-1} \frac{\partial^2 M}{\partial \mu^2} \right)
 - {\rm tr} \left( M^{-1} \frac{\partial M}{\partial \mu}
                   M^{-1} \frac{\partial M}{\partial \mu} \right) 
\label{eq:dermu2} \\
\frac{\partial^3 \ln \det M}{\partial \mu^3}
 &=& 
{\rm tr} \left( M^{-1} \frac{\partial^3 M}{\partial \mu^3} \right)
 -3 {\rm tr} \left( M^{-1} \frac{\partial M}{\partial \mu}
              M^{-1} \frac{\partial^2 M}{\partial \mu^2} \right)
\nonumber \\ && 
+2 {\rm tr} \left( M^{-1} \frac{\partial M}{\partial \mu}
        M^{-1} \frac{\partial M}{\partial \mu}
        M^{-1} \frac{\partial M}{\partial \mu} \right) 
\label{eq:dermu3} \\
\frac{\partial^4 \ln \det M}{\partial \mu^4}
 &=& 
{\rm tr} \left( M^{-1} \frac{\partial^4 M}{\partial \mu^4} \right)
 -4 {\rm tr} \left( M^{-1} \frac{\partial M}{\partial \mu}
              M^{-1} \frac{\partial^3 M}{\partial \mu^3} \right) \nonumber \\
&& 
-3 {\rm tr} \left( M^{-1} \frac{\partial^2 M}{\partial \mu^2}
        M^{-1} \frac{\partial^2 M}{\partial \mu^2} \right)
 +12 {\rm tr} \left( M^{-1} \frac{\partial M}{\partial \mu}
        M^{-1} \frac{\partial M}{\partial \mu}
        M^{-1} \frac{\partial^2 M}{\partial \mu^2} \right) \nonumber \\
&& 
-6 {\rm tr} \left( M^{-1} \frac{\partial M}{\partial \mu}
        M^{-1} \frac{\partial M}{\partial \mu}
        M^{-1} \frac{\partial M}{\partial \mu}
        M^{-1} \frac{\partial M}{\partial \mu} \right) 
\label{eq:dermu4} \\
\frac{\partial^5 \ln \det M}{\partial \mu^5}
 &=& 
{\rm tr} \left( M^{-1} \frac{\partial^5 M}{\partial \mu^5} \right)
 -5 {\rm tr} \left( M^{-1} \frac{\partial M}{\partial \mu}
              M^{-1} \frac{\partial^4 M}{\partial \mu^4} \right) \nonumber \\
&&
 -10 {\rm tr} \left( M^{-1} \frac{\partial^2 M}{\partial \mu^2}
        M^{-1} \frac{\partial^3 M}{\partial \mu^3} \right)
 +20 {\rm tr} \left( M^{-1} \frac{\partial M}{\partial \mu}
        M^{-1} \frac{\partial M}{\partial \mu}
        M^{-1} \frac{\partial^3 M}{\partial \mu^3} \right) \nonumber \\
&&
 +30 {\rm tr} \left( M^{-1} \frac{\partial M}{\partial \mu}
        M^{-1} \frac{\partial^2 M}{\partial \mu^2}
        M^{-1} \frac{\partial^2 M}{\partial \mu^2} \right) \nonumber \\
&&
 -60 {\rm tr} \left( M^{-1} \frac{\partial M}{\partial \mu}
        M^{-1} \frac{\partial M}{\partial \mu}
        M^{-1} \frac{\partial M}{\partial \mu}
        M^{-1} \frac{\partial^2 M}{\partial \mu^2} \right) \nonumber \\
&&
 +24 {\rm tr} \left( M^{-1} \frac{\partial M}{\partial \mu}
        M^{-1} \frac{\partial M}{\partial \mu}
        M^{-1} \frac{\partial M}{\partial \mu}
        M^{-1} \frac{\partial M}{\partial \mu}
        M^{-1} \frac{\partial M}{\partial \mu} \right) 
\label{eq:dermu5} \\
\frac{\partial^6 \ln \det M}{\partial \mu^6}
 &=& 
{\rm tr} \left( M^{-1} \frac{\partial^6 M}{\partial \mu^6} \right)
 -6 {\rm tr} \left( M^{-1} \frac{\partial M}{\partial \mu}
              M^{-1} \frac{\partial^5 M}{\partial \mu^5} \right) \nonumber \\
&&
 -15 {\rm tr} \left( M^{-1} \frac{\partial^2 M}{\partial \mu^2}
        M^{-1} \frac{\partial^4 M}{\partial \mu^4} \right)
 -10 {\rm tr} \left( M^{-1} \frac{\partial^3 M}{\partial \mu^3}
        M^{-1} \frac{\partial^3 M}{\partial \mu^3} \right) \nonumber \\
&&
 +30 {\rm tr} \left( M^{-1} \frac{\partial M}{\partial \mu}
        M^{-1} \frac{\partial M}{\partial \mu}
        M^{-1} \frac{\partial^4 M}{\partial \mu^4} \right)
 +60 {\rm tr} \left( M^{-1} \frac{\partial M}{\partial \mu}
        M^{-1} \frac{\partial^2 M}{\partial \mu^2}
        M^{-1} \frac{\partial^3 M}{\partial \mu^3} \right) \nonumber \\
&&
 +60 {\rm tr} \left( M^{-1} \frac{\partial^2 M}{\partial \mu^2}
        M^{-1} \frac{\partial M}{\partial \mu}
        M^{-1} \frac{\partial^3 M}{\partial \mu^3} \right)
 +30 {\rm tr} \left( M^{-1} \frac{\partial^2 M}{\partial \mu^2}
        M^{-1} \frac{\partial^2 M}{\partial \mu^2}
        M^{-1} \frac{\partial^2 M}{\partial \mu^2} \right) \nonumber \\
&&
 -120 {\rm tr} \left( M^{-1} \frac{\partial M}{\partial \mu}
        M^{-1} \frac{\partial M}{\partial \mu}
        M^{-1} \frac{\partial M}{\partial \mu}
        M^{-1} \frac{\partial^3 M}{\partial \mu^3} \right) \nonumber \\
&&
 -180 {\rm tr} \left( M^{-1} \frac{\partial M}{\partial \mu}
        M^{-1} \frac{\partial M}{\partial \mu}
        M^{-1} \frac{\partial^2 M}{\partial \mu^2}
        M^{-1} \frac{\partial^2 M}{\partial \mu^2} \right) \nonumber \\
&&
 -90 {\rm tr} \left( M^{-1} \frac{\partial M}{\partial \mu}
        M^{-1} \frac{\partial^2 M}{\partial \mu^2}
        M^{-1} \frac{\partial M}{\partial \mu}
        M^{-1} \frac{\partial^2 M}{\partial \mu^2} \right) \nonumber \\
&&
 +360 {\rm tr} \left( M^{-1} \frac{\partial M}{\partial \mu}
        M^{-1} \frac{\partial M}{\partial \mu}
        M^{-1} \frac{\partial M}{\partial \mu}
        M^{-1} \frac{\partial M}{\partial \mu}
        M^{-1} \frac{\partial^2 M}{\partial \mu^2} \right) \nonumber \\
&&
 -120 {\rm tr} \left( M^{-1} \frac{\partial M}{\partial \mu}
        M^{-1} \frac{\partial M}{\partial \mu}
        M^{-1} \frac{\partial M}{\partial \mu}
        M^{-1} \frac{\partial M}{\partial \mu}
        M^{-1} \frac{\partial M}{\partial \mu}
        M^{-1} \frac{\partial M}{\partial \mu} \right) .
\label{eq:dermu6}
\end{eqnarray}
\end{small}

\section{Algebraic techniques for the interaction measure}
Eq.~(13) for the coefficients $b_{nm}(T)$ contains three types of 
derivatives of the fermion matrix with respect to the chemical potentials. We tackle them
separately in the following.
\subsection{First type of derivative}
Here we give the method \cite{Allton:2005gk} for calculating the derivative
\be
 \left.{\partial^{n+m} \langle M_f^{-1}\rangle\over \partial(\mu_l N_t)^n
\partial(\mu_h N_t)^m}\right|_{\mu_{l,h}=0}.
\ee
A convenient place to start in this case is by defining the observables 
\bea
\B_{nm} &\equiv&\left\langle {\rm e}^{-L_{0}}{\rm e}^{-H_{0}}\frac{\partial^n ({\rm tr}\, M_l^{-1}{\rm e}^{L_{0}})}{\partial\mu_l^n}
\frac{\partial^m {\rm e}^{H_{0}}}{\partial\mu_h^m}\right\rangle\\
\B^\prime_{nm} &\equiv&\left\langle {\rm e}^{-L_{0}}{\rm e}^{-H_{0}}\frac{\partial^n {\rm e}^{L_{0}}}{\partial\mu_l^n}
\frac{\partial^m ({\rm tr}\, M_h^{-1}{\rm e}^{H_{0}})}{\partial\mu_h^m}\right\rangle.
\eea 
The above means
\bea
\B_{00}&\equiv&\left\langle{\rm tr}\, M_l^{-1}\right\rangle\\
\B^\prime_{00}&\equiv&\left\langle{\rm tr}\, M_h^{-1}\right\rangle.
\eea
It follows that
\bea
\frac{\partial \B_{nm}}{\partial \mu_l}& =& \B_{n+1,m} - \A_{10}\B_{nm}\\
\frac{\partial \B_{nm}}{\partial \mu_h}& =& \B_{n,m+1} - \A_{01}\B_{nm}\\
\frac{\partial \B^\prime_{nm}}{\partial \mu_l}& =& \B^\prime_{n+1,m} - \A_{10}\B^\prime_{nm}\\
\frac{\partial \B^\prime_{nm}}{\partial \mu_h}& =& \B^\prime_{n,m+1} - \A_{01}\B^\prime_{nm}.
\eea
Using the above and then applying $\mu_{l,h}=0$ we get
\bea
%
\frac{\partial^2\left\langle {\rm tr}M_l^{-1} \right\rangle}{\partial \mu_l^2} & =&
\B_{20}-\A_{20}\B_{00}\nn\\
%
%
\frac{\partial^4\left\langle {\rm tr}M_l^{-1} \right\rangle}{\partial \mu_l^4} & =&
\B_{40}-6\A_{20}\B_{20} +6\A_{20}^2\B_{00}-\A_{40}\B_{00}\nn\\
%
%
\frac{\partial^6\left\langle {\rm tr}M_l^{-1} \right\rangle}{\partial \mu_l^6} & =&
\B_{60} - \A_{60}\B_{00} -15\A_{20}\B_{40}-15\A_{40}\B_{20} +30\A_{20}\A_{40}\B_{00}
+ 90\A_{20}^2\B_{20} -90\A_{20}^3\B_{00}\nn\\
%
\frac{\partial^2\left\langle {\rm tr}M_l^{-1} \right\rangle}{\partial \mu_h^2} & =&
\B_{02}-\A_{02}\B_{00}\nn\\
%
%
\frac{\partial^4\left\langle {\rm tr}M_l^{-1} \right\rangle}{\partial \mu_h^4} & =&
\B_{04}-6\A_{02}\B_{02} +6\A_{02}^2\B_{00}-\A_{04}\B_{00}\nn\\
%
%
\frac{\partial^6\left\langle {\rm tr}M_l^{-1} \right\rangle}{\partial \mu_h^6} & =&
\B_{06} - \A_{06}\B_{00} -15\A_{02}\B_{04}-15\A_{04}\B_{02} +30\A_{02}\A_{04}\B_{00}
+ 90\A_{02}^2\B_{02} -90\A_{02}^3\B_{00}\nn\\
%
\frac{\partial^2\left\langle {\rm tr}M_l^{-1} \right\rangle}{\partial \mu_l\partial \mu_h} & =&
\B_{11} -\A_{11}\B_{00}\nn\\
\frac{\partial^4\left\langle {\rm tr}M_l^{-1} \right\rangle}{\partial \mu_l^2\partial \mu_h^2} & =&
\B_{22} - \A_{22}\B_{00} +2\B_{00}\A_{02}\A_{20}+4\B_{00}\A_{11}^2 -4\B_{11}\A_{11} -\A_{02}\B_{20} -\A_{20}\B_{02}\nn\\
\frac{\partial^4\left\langle {\rm tr}M_l^{-1} \right\rangle}{\partial \mu_l^1\partial \mu_h^3} & =&
\B_{13} - \A_{13}\B_{00}-3\A_{02}\B_{11} - 3\A_{11}\B_{02} +6\A_{11}\A_{02}\B_{00}\nn\\
\frac{\partial^4\left\langle {\rm tr}M_l^{-1} \right\rangle}{\partial \mu_l^3\partial \mu_h^1} & =&
\B_{31} - \A_{31}\B_{00}-3\A_{20}\B_{11} - 3\A_{11}\B_{20} +6\A_{11}\A_{20}\B_{00}\nn
\eea
\bea
\frac{\partial^6\left\langle {\rm tr}M_l^{-1} \right\rangle}{\partial \mu_l^2\partial \mu_h^4} & =&
\B_{24} -\A_{24}\B_{00} -8\A_{11}\B_{13} +16\A_{13}\A_{11}\B_{00}-8\A_{13}\B_{11} -\A_{04}\B_{20} -\A_{20}\B_{04} \nn\\
&& -6\A_{02}\B_{22} +6\A_{02}^2\B_{20} -6\A_{22}\B_{02} +48 \A_{02}\A_{11}\B_{11} +12\A_{22}\A_{02}\B_{00} +12 \A_{20}\A_{02}\B_{02}\nn\\
&& +24\A_{11}^2\B_{02} -72\A_{11}^2\A_{02}\B_{00} -18\A_{20}\A_{02}^2\B_{00} +2\A_{20}\A_{04}\B_{00}\nn\\
\frac{\partial^6\left\langle {\rm tr}M_l^{-1} \right\rangle}{\partial \mu_l^4\partial \mu_h^2} & =&
\B_{42} -\A_{42}\B_{00} -8\A_{11}\B_{31} +16\A_{31}\A_{11}\B_{00}-8\A_{31}\B_{11} -\A_{40}\B_{02} -\A_{02}\B_{40} \nn\\
&& -6\A_{20}\B_{22} +6\A_{20}^2\B_{02} -6\A_{22}\B_{20} +48 \A_{20}\A_{11}\B_{11} +12\A_{22}\A_{20}\B_{00} +12 \A_{02}\A_{20}\B_{20}\nn\\
&& +24\A_{11}^2\B_{20} -72\A_{11}^2\A_{20}\B_{00} -18\A_{02}\A_{20}^2\B_{00} +2\A_{02}\A_{40}\B_{00}\nn\\
\frac{\partial^6\left\langle {\rm tr}M_l^{-1} \right\rangle}{\partial \mu_l^1\partial \mu_h^5} & =&
\B_{15} -\A_{15}\B_{00} -10\A_{02}\B_{13} +20\A_{13}\A_{02}\B_{00} -5\A_{04}\B_{11} -10\A_{13}\B_{02} -5\A_{11}\B_{04}\nn\\
&& + 30 \A_{02}^2\B_{11} +60\A_{11}\A_{02}\B_{02} +10\A_{11}\A_{04}\B_{00}-90\A_{11}\A_{02}^2\B_{00}\nn\\
\frac{\partial^6\left\langle {\rm tr}M_l^{-1} \right\rangle}{\partial \mu_l^5\partial \mu_h^1} & =&
\B_{51} -\A_{51}\B_{00} -10\A_{20}\B_{31} +20\A_{31}\A_{20}\B_{00} -5\A_{40}\B_{11} -10\A_{31}\B_{20} -5\A_{11}\B_{40}\nn\\
&& + 30 \A_{20}^2\B_{11} +60\A_{11}\A_{20}\B_{20} +10\A_{11}\A_{40}\B_{00}-90\A_{11}\A_{20}^2\B_{00}\nn\\
\frac{\partial^6\left\langle {\rm tr}M_l^{-1} \right\rangle}{\partial \mu_l^3\partial \mu_h^3} & =&
\B_{33}-\A_{33}\B_{00} -3\A_{20}\B_{13} -3\A_{02}\B_{31} -9\A_{11}\B_{22} -9 \A_{22}\B_{11} -3 \A_{13}\B_{20} -3\A_{31}\B_{02}\nn\\
&& +36\A_{11}^2\B_{11} -36\A_{11}^3\B_{00} +6\A_{13}\A_{20}\B_{00} +6\A_{31}\A_{02}\B_{00}+ 18\A_{02}\A_{20}\B_{11} \nn\\
&& + 18\A_{11}\A_{22}\B_{00} +18\A_{02}\A_{11}\B_{20} +18\A_{11}\A_{20}\B_{02} -54\A_{11}\A_{02}\A_{20}\B_{00}.\nn
\eea
Replacing $\B$ with $\B^\prime$ in the above  we get 
the expressions for the derivatives of 
$\left\langle {\rm tr}M_h^{-1} \right\rangle$.
Let 
\bea
{\el}_n &=& \frac{\partial^n {\rm tr}\,M_l^{-1}}{\partial \mu_l^n}\\
\h_n &=& \frac{\partial^n {\rm tr}\,M_h^{-1}}{\partial \mu_h^n},
\eea
then explicitly we have
\bea
\B_{00}& =& \left\langle l_0 \right\rangle = \left\langle {\rm tr}M_l^{-1} \right\rangle\nn\\
\B_{10} &=&
 \left\langle l_1  \right\rangle
+\left\langle l_0L_1  \right\rangle\nn\\
\B_{20} &=&
 \left\langle l_2  \right\rangle
+ 2\left\langle l_1 L_1  \right\rangle
+ \left\langle l_0 L_2 \right\rangle
+ \left\langle l_0L_1^2 \right\rangle\nn\\
\B_{30} &=&
 \left\langle l_3 \right\rangle
+3 \left\langle l_2L_1 \right\rangle
+ 3\left\langle l_1L_2 \right\rangle
+ \left\langle l_0L_3  \right\rangle
+ 3\left\langle l_1L_1^2  \right\rangle
+ 3\left\langle l_0L_1L_2 \right\rangle
+ \left\langle l_0L_1^3 \right\rangle\nn\\
\B_{40} &=&
\left\langle l_4 \right\rangle
+4\left\langle l_3L_1\right\rangle
+6\left\langle l_2L_2\right\rangle
+4\left\langle l_1L_3 \right\rangle
+\left\langle l_0L_4 \right\rangle
+6\left\langle l_2L_1^2\right\rangle
+12\left\langle l_1L_1L_2\right\rangle
+3\left\langle l_0 L_2^2 \right\rangle\nn\\
&&+4\left\langle l_0L_1L_3\right\rangle
+4\left\langle l_1L_1^3\right\rangle
+6\left\langle l_0L_1^2L_2\right\rangle
+\left\langle l_0L_1^4\right\rangle\nn\\
\B_{50}&=&
\left\langle l_5 \right\rangle
+30\left\langle l_2L_1L_2 \right\rangle
+30\left\langle l_1L_1^2L_2\right\rangle
+20\left\langle l_1L_1L_3\right\rangle
+10\left\langle l_0L_1^3L_2\right\rangle
+\left\langle l_0L_5 \right\rangle
+5\left\langle l_1L_1^4\right\rangle
+10\left\langle l_2L_3\right\rangle\nn\\
&&
+5\left\langle l_4L_1\right\rangle
+10\left\langle l_3L_2 \right\rangle
+10\left\langle l_3L_1^2\right\rangle
+5\left\langle l_1L_4 \right\rangle
+15\left\langle l_1L_2^2\right\rangle
+10\left\langle l_2L_1^3\right\rangle
+\left\langle l_0L_1^5\right\rangle\nn\\
&&+10\left\langle l_0L_2L_3 \right\rangle
+5\left\langle l_0L_1L_4\right\rangle
+15\left\langle l_0L_1L_2^2\right\rangle
+10\left\langle l_0L_1^2L_3\right\rangle\nn\\
\B_{60}&=&
\left\langle l_6 \right\rangle
+60 \left\langle l_1L_2L_3 \right\rangle
+15 \left\langle l_0L_1^4L_2\right\rangle
+20 \left\langle l_0L_1^3L_3\right\rangle
+90 \left\langle l_2L_1^2L_2\right\rangle
+90 \left\langle l_1L_1L_2^2\right\rangle
+ \left\langle l_0L_6\right\rangle\nn\\
&&
+6 \left\langle l_1L_1^5\right\rangle
+45 \left\langle l_2L_2^2\right\rangle
+15 \left\langle l_2L_1^4\right\rangle
+20 \left\langle l_3L_3\right\rangle
+15 \left\langle l_2L_4\right\rangle
+6 \left\langle l_5L_1\right\rangle
+15 \left\langle l_4L_2\right\rangle
+15 \left\langle l_4L_1^2\right\rangle\nn\\
&&
+20 \left\langle l_3L_1^3\right\rangle
+ 10\left\langle l_0L_3^2\right\rangle
+ 15\left\langle l_0L_2^3\right\rangle
+6 \left\langle l_1L_5\right\rangle
+60 \left\langle l_3L_1L_2\right\rangle
+60 \left\langle l_2L_1L_3\right\rangle
+60 \left\langle l_1L_1^2L_3\right\rangle\nn\\
&&
+30 \left\langle l_1L_1L_4\right\rangle
+ 60\left\langle l_1L_1^3L_2\right\rangle
+ 45\left\langle l_0L_1^2L_2^2 \right\rangle
+15 \left\langle l_0L_2L_4\right\rangle
+ 6\left\langle l_0L_1L_5\right\rangle
+ 15\left\langle l_0L_1^2L_4\right\rangle\nn\\
&&
+ 60\left\langle l_0L_1L_2L_3\right\rangle
+ \left\langle l_0L_1^6\right\rangle\nn\\
{\cal B}_{02}  &=& 
\left\langle l_0H_2 \right\rangle
+\left\langle l_0H_1^2 \right\rangle
\nn\\
{\cal B}_{04}  &=& 
\left\langle l_0 H_4 \right\rangle
+4\left\langle l_0H_3 H_1 \right\rangle
+3\left\langle l_0H_2^2 \right\rangle
+6\left\langle l_0H_2 H_1^2 \right\rangle
+\left\langle l_0H_1^4 \right\rangle
\nn\\
{\cal B}_{06}  &=& 
\left\langle l_0H_6 \right\rangle
+6\left\langle l_0H_5 H_1 \right\rangle
+15\left\langle l_0H_4 H_2 \right\rangle
+10\left\langle l_0H_3^2 \right\rangle
+15\left\langle l_0H_4 H_1^2 \right\rangle
+60\left\langle l_0H_3 H_2 H_1 \right\rangle
\nonumber \\ && 
+15\left\langle l_0H_2^3 \right\rangle
+20\left\langle l_0H_3 H_1^3 \right\rangle
+45\left\langle l_0H_2^2 H_1^2 \right\rangle
+15\left\langle l_0H_2 H_1^4 \right\rangle
+\left\langle l_0H_1^6 \right\rangle  \quad \nn\\
{\cal B}_{11}  &=&
\left\langle l_1 H_1 \right\rangle
+\left\langle l_0L_1H_1  \right\rangle\nn\\
{\cal B}_{22}  &=&
 \left\langle (l_2  
+ 2 l_1 L_1  
+  l_0 L_2 
+  l_0L_1^2)(H_2+H_1^2) \right\rangle\nn\\
{\cal B}_{31}  &=&
 \left\langle l_3H_1 \right\rangle
+3 \left\langle l_2L_1H_1 \right\rangle
+ 3\left\langle l_1L_2H_1 \right\rangle
+ \left\langle l_0L_3H_1  \right\rangle
+ 3\left\langle l_1L_1^2H_1  \right\rangle
+ 3\left\langle l_0L_1L_2H_1 \right\rangle
+ \left\langle l_0L_1^3H_1 \right\rangle\nn\\
{\cal B}_{13}  &=&
 \left\langle (l_1 
+ l_0L_1)
(H_3+3H_2H_1 +H_1^3) \right\rangle\nn\\
{\cal B}_{42}  &=&
\left\langle (l_4 
+4 l_3L_1
+6 l_2L_2
+4 l_1L_3 
+ l_0L_4 
+6 l_2L_1^2
+12 l_1L_1L_2
+3 l_0 L_2^2 \right.\nn\\
&&\left.+4 l_0L_1L_3
+4 l_1L_1^3
+6 l_0L_1^2L_2
+ l_0L_1^4)
(H_2+H_1^2)
\right\rangle\nn\\
{\cal B}_{24}  &=&
 \left\langle (l_2  
+ 2 l_1 L_1  
+  l_0 L_2 
+  l_0L_1^2)(H_4+4H_3H_1+3H_2^2 +6H_2H_1^2 +H_1^4) \right\rangle\nn\\
{\cal B}_{51}  &=&
\left\langle l_5 H_1\right\rangle
+30\left\langle l_2L_1L_2H_1 \right\rangle
+30\left\langle l_1L_1^2L_2H_1\right\rangle
+20\left\langle l_1L_1L_3H_1\right\rangle
+10\left\langle l_0L_1^3L_2H_1\right\rangle
+\left\langle l_0L_5H_1 \right\rangle\nn\\
&&
+5\left\langle l_1L_1^4H_1\right\rangle
+10\left\langle l_2L_3H_1\right\rangle
+5\left\langle l_4L_1H_1\right\rangle
+10\left\langle l_3L_2H_1 \right\rangle
+10\left\langle l_3L_1^2H_1\right\rangle
+5\left\langle l_1L_4H_1 \right\rangle\nn\\
&&
+15\left\langle l_1L_2^2H_1\right\rangle
+10\left\langle l_2L_1^3H_1\right\rangle
+\left\langle l_0L_1^5H_1\right\rangle
+10\left\langle l_0L_2L_3H_1 \right\rangle
+5\left\langle l_0L_1L_4H_1\right\rangle
+15\left\langle l_0L_1L_2^2H_1\right\rangle\nn\\
&&
+10\left\langle l_0L_1^2L_3H_1\right\rangle\nn\\
{\cal B}_{15}  &=&\left\langle
(l_1 
+ l_0L_1)(H_5+5H_4H_1 +10H_3H_2 +10H_3H_1^2 +15H_2^2H_1+10H_2H_1^3+H_1^5)\right\rangle\nn\\
{\cal B}_{33}  &=&\left\langle
(l_3 
+3  l_2L_1 
+ 3 l_1L_2 
+  l_0L_3  
+ 3 l_1L_1^2  
+ 3 l_0L_1L_2 
+  l_0L_1^3)
(H_3+3H_2H_1 +H_1^3)
\right\rangle.\nn
\eea
From the above expressions it is easy to get the $\B^\prime$ expressions by substitutions
\be
\B^\prime_{mn}=\B_{nm}(l\rightarrow h, L\leftrightarrow H).
\ee 
Explicitly $l_n$ and $h_n$ are the derivatives below with $M=M_{l,h}$ and $\mu=\mu_{l,h}$.
\begin{small}
\begin{eqnarray}
\frac{\partial {\rm tr} M^{-1}}{\partial \mu}
 &=& 
- {\rm tr} \left( M^{-1} \frac{\partial M}{\partial \mu}
 M^{-1} \right) \\
\frac{\partial^2 {\rm tr} M^{-1}}{\partial \mu^2}
 &=& 
- {\rm tr} \left( M^{-1} \frac{\partial^2 M}{\partial \mu^2}
 M^{-1} \right)
 + 2 {\rm tr} \left( M^{-1} \frac{\partial M}{\partial \mu}
    M^{-1} \frac{\partial M}{\partial \mu} M^{-1} \right) \\
\frac{\partial^3 {\rm tr} M^{-1}}{\partial \mu^3}
 &=& 
- {\rm tr} \left( M^{-1} \frac{\partial^3 M}{\partial \mu^3}
 M^{-1} \right)
 +3 {\rm tr} \left( M^{-1} \frac{\partial^2 M}{\partial \mu^2}
    M^{-1} \frac{\partial M}{\partial \mu} M^{-1} \right)  \\
&& 
+3 {\rm tr} \left( M^{-1} \frac{\partial M}{\partial \mu}
    M^{-1} \frac{\partial^2 M}{\partial \mu^2} M^{-1} \right)
-6 {\rm tr} \left( M^{-1} \frac{\partial M}{\partial \mu}
    M^{-1} \frac{\partial M}{\partial \mu} M^{-1}
    \frac{\partial M}{\partial \mu} M^{-1} \right) \nonumber \\
\frac{\partial^4 {\rm tr} M^{-1}}{\partial \mu^4}
 &=& 
- {\rm tr} \left( M^{-1} \frac{\partial^4 M}{\partial \mu^4}
 M^{-1} \right)
 +4 {\rm tr} \left( M^{-1} \frac{\partial^3 M}{\partial \mu^3}
    M^{-1} \frac{\partial M}{\partial \mu} M^{-1} \right)  \\
&& 
+6 {\rm tr} \left( M^{-1} \frac{\partial^2 M}{\partial \mu^2}
    M^{-1} \frac{\partial^2 M}{\partial \mu^2} M^{-1} \right)
+4 {\rm tr} \left( M^{-1} \frac{\partial M}{\partial \mu}
    M^{-1} \frac{\partial^3 M}{\partial \mu^3} M^{-1} \right) \nonumber \\
&& 
-12 {\rm tr} \left( M^{-1} \frac{\partial^2 M}{\partial \mu^2}
    M^{-1} \frac{\partial M}{\partial \mu} M^{-1}
    \frac{\partial M}{\partial \mu} M^{-1} \right) \nonumber \\
&& 
-12 {\rm tr} \left( M^{-1} \frac{\partial M}{\partial \mu}
    M^{-1} \frac{\partial^2 M}{\partial \mu^2} M^{-1}
    \frac{\partial M}{\partial \mu} M^{-1} \right) \nonumber \\
&& 
-12 {\rm tr} \left( M^{-1} \frac{\partial M}{\partial \mu}
    M^{-1} \frac{\partial M}{\partial \mu} M^{-1}
    \frac{\partial^2 M}{\partial \mu^2} M^{-1} \right) \nonumber \\
&& 
+24 {\rm tr} \left( M^{-1} \frac{\partial M}{\partial \mu}
    M^{-1} \frac{\partial M}{\partial \mu} M^{-1}
    \frac{\partial M}{\partial \mu} M^{-1}
    \frac{\partial M}{\partial \mu} M^{-1} \right) \nonumber\\
\frac{\partial^5 {\rm tr} M^{-1}}{\partial \mu^5}
 &=&
-{\rm tr} \left(\Mfi M^{-1}\right)
+5{\rm tr} \left(\Mf\Mo  M^{-1}\right)\\
&&+5{\rm tr} \left(\Mo\Mf M^{-1}\right)
+10{\rm tr} \left(\Mt\Mth M^{-1}\right)\nn\\
&&+10{\rm tr} \left(\Mth\Mt M^{-1}\right)
-30{\rm tr} \left(\Mo\Mt\Mt M^{-1}\right)\nn\\
&&-30{\rm tr} \left(\Mt\Mt\Mo M^{-1}\right)
-30{\rm tr} \left(\Mt\Mo\Mt M^{-1}\right)\nn\\
&&-20{\rm tr} \left(\Mo\Mo\Mth M^{-1}\right)
-20{\rm tr} \left(\Mo\Mth\Mo M^{-1}\right)\nn\\
&&-20{\rm tr} \left(\Mth\Mo\Mo M^{-1}\right)
+60{\rm tr} \left(\Mt\Mo\Mo\Mo M^{-1}\right)\nn\\
&&+60{\rm tr} \left(\Mo\Mt\Mo\Mo M^{-1}\right)
+60{\rm tr} \left(\Mo\Mo\Mt\Mo M^{-1}\right)\nn\\
&&
+60{\rm tr} \left(\Mo\Mo\Mo\Mt M^{-1}\right)\nn\\
&&
-120{\rm tr} \left(\Mo\Mo\Mo\Mo\Mo M^{-1}\right) \nn\\
\frac{\partial^6 {\rm tr} M^{-1}}{\partial \mu^6}
 &=&
-{\rm tr} \left(\Ms M^{-1}\right)
+6{\rm tr} \left(\Mo\Mfi M^{-1}\right)\\
&&+6{\rm tr} \left(\Mfi\Mo M^{-1}\right)
+15{\rm tr} \left(\Mt\Mf M^{-1}\right)\nn\\
&&+15{\rm tr} \left(\Mf\Mt M^{-1}\right)
+20{\rm tr} \left(\Mth\Mth M^{-1}\right)\nn\\
&&-30{\rm tr} \left(\Mo\Mo\Mf M^{-1}\right)
-30{\rm tr} \left(\Mo\Mf\Mo M^{-1}\right)\nn\\
&&-30{\rm tr} \left(\Mf\Mo\Mo M^{-1}\right)
-60{\rm tr} \left(\Mo\Mt\Mth M^{-1}\right)\nn\\
&&-60{\rm tr} \left(\Mo\Mth\Mt M^{-1}\right)
-60{\rm tr} \left(\Mth\Mt\Mo M^{-1}\right)\nn\\
&&-60{\rm tr} \left(\Mt\Mth\Mo M^{-1}\right)
-60{\rm tr} \left(\Mt\Mo\Mth M^{-1}\right)\nn\\
&& -90{\rm tr} \left(\Mt\Mt\Mt M^{-1}\right)
-60{\rm tr} \left(\Mth\Mo\Mt M^{-1}\right)\nn\\
&&+120{\rm tr} \left(\Mth\Mo\Mo\Mo M^{-1}\right)\nn\\
&&+120{\rm tr} \left(\Mo\Mth\Mo\Mo M^{-1}\right)\nn\\
&&+120{\rm tr} \left(\Mo\Mo\Mth\Mo M^{-1}\right)\nn\\
&&+120{\rm tr} \left(\Mo\Mo\Mo\Mth M^{-1}\right)\nn\\
&&+180{\rm tr} \left(\Mt\Mt\Mo\Mo M^{-1}\right)\nn\\
&&+180{\rm tr} \left(\Mt\Mo\Mt\Mo M^{-1}\right)\nn\\
&&+180{\rm tr} \left(\Mt\Mo\Mo\Mt M^{-1}\right)\nn\\
&&+180{\rm tr} \left(\Mo\Mt\Mt\Mo M^{-1}\right)\nn\\
&&+180{\rm tr} \left(\Mo\Mt\Mo\Mt M^{-1}\right)\nn\\
&&+180{\rm tr} \left(\Mo\Mo\Mt\Mt M^{-1}\right)\nn\\
&&-360{\rm tr} \left(\Mt\Mo\Mo\Mo\Mo M^{-1}\right)\nn\\
&&-360{\rm tr} \left(\Mo\Mt\Mo\Mo\Mo M^{-1}\right)\nn\\
&&-360{\rm tr} \left(\Mo\Mo\Mt\Mo\Mo M^{-1}\right)\nn\\
&&-360{\rm tr} \left(\Mo\Mo\Mo\Mt\Mo M^{-1}\right)\nn\\
&&-360{\rm tr} \left(\Mo\Mo\Mo\Mo\Mt M^{-1}\right)\nn\\
&&+720{\rm tr} \left(\Mo\Mo\Mo\Mo\Mo\Mo M^{-1}\right).\nn
\end{eqnarray}
\end{small}

\subsection{Second type of derivative}
The next term we are concerned with is the derivative
\be
\left.{\partial^{n+m}
\langle M_f^{-1}\frac{dM_f}{du_0}\rangle
\over \partial(\mu_l N_t)^n\partial(\mu_h N_t)^m}\right|_{\mu_{l,h}=0}.
\ee
Here we start start from the definitions
\bea
\C_{nm} &\equiv&\left\langle {\rm e}^{-L_{0}}{\rm e}^{-H_{0}}\frac{\partial^n [{\rm tr}\, (M_l^{-1}\frac{dM_l}{du_0}){\rm e}^{L_{0}}]}{\partial\mu_l^n}
\frac{\partial^m {\rm e}^{H_{0}}}{\partial\mu_h^m}\right\rangle\\
\C^\prime_{nm} &\equiv&\left\langle {\rm e}^{-L_{0}}{\rm e}^{-H_{0}}\frac{\partial^n {\rm e}^{L_{0}}}{\partial\mu_l^n}
\frac{\partial^m [{\rm tr}\, (M_h^{-1}\frac{dM_h}{du_0}){\rm e}^{H_{0}}]}{\partial\mu_h^m}\right\rangle.
\eea
From the above 
\bea
\C_{00}&\equiv&\left\langle{\rm tr}\, (M_l^{-1}\frac{dM_l}{du_0})\right\rangle\\
\C^\prime_{00}&\equiv&\left\langle{\rm tr}\,( M_h^{-1}\frac{dM_h}{du_0})\right\rangle.
\eea
The following can be proven true
\bea
\frac{\partial \C_{nm}}{\partial \mu_l}& =& \C_{n+1,m} - \A_{10}\C_{nm}\\
\frac{\partial \C_{nm}}{\partial \mu_h}& =& \C_{n,m+1} - \A_{01}\C_{nm}\\
\frac{\partial \C^\prime_{nm}}{\partial \mu_l}& =& \C^\prime_{n+1,m} - \A_{10}\C^\prime_{nm}\\
\frac{\partial \C^\prime_{nm}}{\partial \mu_h}& =& \C^\prime_{n,m+1} - \A_{01}\C^\prime_{nm}.
\eea
The derivatives 
\be
\frac{\partial^n \left\langle{\rm tr}\, (M_{l,h}^{-1}\frac{dM_{l,h}}{du_0})\right\rangle}{\partial\mu_{l,h}^n}
\ee
have the form of the derivatives of $\left\langle{\rm tr}\, (M_{l,h}^{-1})\right\rangle$ in the previous section with the substitutions
$\B_{nm}\rightarrow \C_{nm}$ and $\B_{nm}^\prime\rightarrow \C_{nm}^\prime$.
The explicit forms of $\C_{nm}$ and $\C_{nm}^\prime$ are the same as for $\B_{nm}$ and $\B_{nm}^\prime$ with the substitutions
$l_n\rightarrow\lambda_{n}$ and $h_n\rightarrow\chi_n$, where
\bea
\lambda_n &=& \frac{\partial^n {\rm tr}\, (M_{l}^{-1}\frac{dM_{h}}{du_0})}{\partial\mu_{l}^n}\\
\chi_n &=& \frac{\partial^n {\rm tr}\, (M_{h}^{-1}\frac{dM_{h}}{du_0})}{\partial\mu_{h}^n}.
\eea

These derivatives have the form below with $M=M_{l,h}$ and $\mu=\mu_{l,h}$:
\begin{small}
\bea
\frac{\partial {\rm tr}\, (M^{-1}\frac{dM}{du_0})}{\partial\mu} &=&{\rm tr}\,\left( \mo\fz +\mz\fo \right)\nn\\
\frac{\partial^2 {\rm tr}\, (M^{-1}\frac{dM}{du_0})}{\partial\mu^2} &=&
{\rm tr}\,\left(
\mt\fz+2\mo\fo+\mz\ft
\right)\nn\\
\frac{\partial^3 {\rm tr}\, (M^{-1}\frac{dM}{du_0})}{\partial\mu^3} &=&
{\rm tr}\,\left(
\mth\fz+3\mt\fo+3\mo\ft+\mz\fth
\right)\nn\\
\frac{\partial^4 {\rm tr}\, (M^{-1}\frac{dM}{du_0})}{\partial\mu^4} &=&
{\rm tr}\,\left(
\mf\fz+4\mth\fo+6\mt\ft+4\mo\fth\nn\right.\\
&&+\left.\mz\ff
\right)\nn\\
\frac{\partial^5 {\rm tr}\, (M^{-1}\frac{dM}{du_0})}{\partial\mu^5} &=&
{\rm tr}\,\left(
\mfi\fz+5\mf\fo+5\mo\ff+10\mth\ft\nn\right.\\
&&\left.+10\mt\fth+\mz\ffi
\right)\nn\\
\frac{\partial^6 {\rm tr}\, (M^{-1}\frac{dM}{du_0})}{\partial\mu^6} &=&
{\rm tr}\,\left(
\ms\fz+6\mfi\fo+6\mo\ffi+15\mf\ft\nn\right.\\
&&\left.+15\mt\ff+20\mth\fth+\mz\fs
\right).\nn
\eea
\end{small}
In the above the derivatives of $M^{-1}$ can be taken from the previous 
subsection. The derivative of $M$ with respect both to
the chemical potential and the tadpole factor for the asqtad action, is
\bea
\frac{\partial^n}{\partial\mu^n}\frac{dM}{du_0} &=& 
\frac{1}{2} \eta_0(x) \left[ \frac{d\,U_0^{(F)}(x)}{d\,u_o} {\rm e}^\mu \delta_{x+\hat 0,y} -
 (-1)^n\frac{d\,U_0^{(F)\dagger}(x-\hat{0})}{d\,u_o} {\rm e}^{-\mu} \delta_{x,y+\hat 0} \right. \\\nn 
 &&\left. + (3)^n\frac{d\,U_0^{(L)}(x)}{d\,u_o} {\rm e}^{3\mu} \delta_{x+3\hat 0,y} -
 (-3)^n\frac{d\,U_0^{(L)\dagger}(x-3\hat{0})}{d\,u_o} {\rm e}^{-3\mu} \delta_{x,y+3\hat 0} \right].
\eea

\subsection{Third type of derivative}
The third type is the gauge derivative
\be
\left.{\partial^{n+m}\langle {\cal G}\rangle
\over \partial(\mu_l N_t)^n\partial(\mu_h N_t)^m}\right|_{\mu_{l,h}=0}.
\ee
In this case let
\be
G_{nm}\equiv\left\langle {\cal G}\,{\rm e}^{-L_{0}}{\rm e}^{-H_{0}}\frac{\partial^n {\rm e}^{L_{0}}}{\partial\mu_l^n}
\frac{\partial^m {\rm e}^{H_{0}}}{\partial\mu_h^m}\right\rangle,
\ee
and similarly as before
\bea
\frac{\partial G_{nm}}{\partial \mu_l}&=& G_{n+1,m} - \A_{10}G_{nm}\\
\frac{\partial G_{nm}}{\partial \mu_h}&=& G_{n,m+1} - \A_{01}G_{nm},
\eea
with
\be
G_{00} = \langle {\cal G}\rangle.
\ee
This means that the necessary derivatives $\left.{\partial^{n+m}\langle {\cal G}\rangle
\over \partial(\mu_l N_t)^n\partial(\mu_h N_t)^m}\right|_{\mu_{l,h}=0}$ have the same form as
the derivatives $\left.{\partial^{n+m}{\rm tr}\langle M_f^{-1}\rangle\over \partial(\mu_l N_t)^n\partial(\mu_h N_t)^m}\right|_{\mu_{l,h}=0}$ with $\B_{nm}\rightarrow G_{nm}$.
The $G_{nm}$ observables have very similar form to the $\A_{nm}$ observables, 
but with an additional multiplication by ${\cal G}$ inside the ensemble average brackets of
each term in them. For example:
\be
\left.{\partial^2\langle {\cal G}\rangle
\over \partial\mu_l^2}\right|_{\mu_{l,h}=0} = G_{20} - \A_{20}G_{00}
\ee
and
\be
G_{20} = \langle {\cal G} L_2\rangle + \langle {\cal G} L_1^2\rangle,
\ee
etc.

For example, combining the three types of terms for each flavor, one of the 
simplest of the 
Taylor coefficients in the interaction measure expansion, $b_{20}$, becomes
\bea
b_{20}&=&-{1\over 2!}{N_t\over N_s^3} \left[ 
\left.\frac{1}{2}\frac{d(m_la)}{d\ln a}\right|_{\mu_{l,h}=0}( \B_{20}-\A_{20}\B_{00})
+\frac{1}{2}\left.\frac{du_0}{d\ln a}\right|_{\mu_{l,h}=0}(\C_{20}-\A_{20}\C_{00})\right.\nn\\
&&+\left. \frac{1}{4}\left.\frac{d(m_ha)}{d\ln a}\right|_{\mu_{l,h}=0}( \B^\prime_{20}-\A_{20}\B^\prime_{00}) 
+\frac{1}{4}\left.\frac{du_0}{d\ln a}\right|_{\mu_{l,h}=0}(\C^\prime_{20}-\A_{20}\C^\prime_{00})\right.\nn\\
&& + \left.G_{20} - \A_{20}G_{00}\right].
\eea

\end{document}